\documentclass[a4paper]{article}
\usepackage{amssymb}
\usepackage{amsmath}
\usepackage{amsthm}

\newcommand{\ssection}[1] {
\refstepcounter{section}
\setcounter{equation}{0}
\setcounter{subsection}{0}
\addcontentsline{toc}{section}{\textbf{\thesection.\ #1}}
\bigskip\bigskip\noindent
\Large
\textbf{\thesection.\ #1}
\nopagebreak\bigskip\nopagebreak
\normalsize
}
\def\thesection{{\arabic{section}}}

\newcommand{\ssubsection}[1]
{
\refstepcounter{subsection}
\addcontentsline{toc}{subsection}{\normalfont\textit{\thesubsection.\ #1}}
\medskip\medskip\noindent
\large\normalfont
\textit{\thesubsection. \ #1}\nopagebreak\bigskip\nopagebreak
\normalsize
}
\def\thesubsection{{{\arabic{section}.\arabic{subsection}}}}

\newtheorem{theorem}{Theorem}[section]
\newtheorem{lemma}[theorem]{Lemma}
\newtheorem{proposition}[theorem]{Proposition}
\newtheorem{corollary}[theorem]{Corollary}

\theoremstyle{definition}

\theoremstyle{remark}
\newtheorem{remark}[theorem]{Remark}

\newcommand{\n}{\mathfrak{n}}
\newcommand{\gc}{\operatorname{gc}}
\newcommand{\hgc}{\hat{\operatorname{gc}}}

\newcommand{\id}{{1 \mskip -5mu {\rm I}}}

\newcommand{\hrpms}{\hat{R}^{(\pm)}_S}
\newcommand{\hrpmxs}{\hat{R}^{(\pm)}_{*,S}}
\newcommand{\rpms}{R^{(\pm)}_S}
\newcommand{\rpmxs}{R^{(\pm)}_{*,S}}
\newcommand{\lpms}{L^{(\pm)}_S}
\newcommand{\lps}{L^{(+)}_S}
\newcommand{\lms}{L^{(-)}_S}
\newcommand{\hrps}{\hat{R}^{(+)}_S}

\newcommand{\hrpxs}{\hat{R}^{(+)}_{*,S}}

\newcommand{\rps}{R^{(+)}_S}
\newcommand{\rms}{R^{(-)}_S}
\newcommand{\rpxs}{R^{(+)}_{*,S}}
\newcommand{\rmxs}{R^{(-)}_{*,S}}
\newcommand{\hrpmsj}{\hat{R}^{(\pm)}_{S,J}}
\newcommand{\hrpmsk}{\hat{R}^{(\pm)}_{S,k}}
\newcommand{\hrpsj}{\hat{R}^{(+)}_{S,J}}

\newcommand{\hrpsk}{\hat{R}^{(+)}_{S,k}}
\newcommand{\hrmsk}{\hat{R}^{(-)}_{S,k}}

\newcommand{\rpmsk}{R^{(\pm)}_{S,k}}

\newcommand{\rpsk}{R^{(+)}_{S,k}}
\newcommand{\rmsk}{R^{(-)}_{S,k}}
\newcommand{\hrpmsjpmk}{\hat{R}^{(\pm)}_{S,J^{\pm}_k}}

\newcommand{\cA}{\mathcal A}

\newcommand{\bC}{\mathbb C}

	\newcommand{\bZ}{\mathbb Z}

	\newcommand{\Z}{\mathbb Z}
\newcommand{\C}{\mathbb C}

\begin{document}



\par\vskip 1cm\vskip 2em

\begin{center}

{\Huge  Subalgebras of $\gc_N$ and Jacobi polynomials}

\par

\vskip 2.5em \lineskip .5em

{\large
\begin{tabular}[t]{c}
$\mbox{Alberto De Sole}\!\!\phantom{m}\mbox{,\ \ Victor G. Kac}$\\
\end{tabular}
\par
}
\medskip
{\small
\begin{tabular}[t]{ll}
& Department of Mathematics, MIT \\
& 77 Massachusetts Avenue, Cambridge, MA 02139, USA \\
& E--mails: {\tt desole@math.mit.edu} \\
& \hspace{1.25cm} {\tt kac@math.mit.edu}
\end{tabular}
}
\bigskip
\end{center}


To Robert Moody on his 60--th birthday.

\vskip 1 em
\centerline{\bf Abstract}
\smallskip
{\small
\noindent
We classify the subalgebras of the general Lie conformal algebra $\gc_N$
that act irreducibly on $\C[\partial]^N$ and that are normalized by the
$\operatorname{sl}_2$--part of a Virasoro element.
The problem turns out to be closely related to classical Jacobi polynomials
$P_n^{(-\sigma,\sigma)}$, $\sigma\in\C$.
The connection goes both ways -- we use in our classification some classical
properties of Jacobi polynomials, and we derive from the theory of conformal
algebras some apparently new properties of Jacobi polynomials.}



\bigskip


\ssection{Introduction: basic definitions and techniques}


\noindent
Lie conformal algebras encode the singular part of the operator
product expansion of formal distributions on the circle.
Therefore they are important for conformal field theory on the one hand,
and for the theory of infinite--dimensional Lie algebras on
the other hand \cite{kac}.


In this paper we study subalgebras of the Lie conformal algebra $gc_N$
associated to the Lie algebra of differential operators on the circle
with coefficients in the space of $N\times N$ matrices over $\bC$.
Quite surprisingly, this problem turns out to be closely related
to classical Jacobi polynomials.


Recall that a \emph{Lie conformal algebra}  $R$ is defined
(see \cite{kac}, \cite{kac3})
as a $\bC[\partial]$-module endowed with a $\bC$-linear map
$R\otimes R \rightarrow \bC[\lambda]\otimes R$ denoted by $a\otimes b
\rightarrow [a \ _\lambda \  b]$
and called $\lambda$-bracket, satisfying the following axioms:
\begin{eqnarray*}
& [\partial a  \ _\lambda \  b] = -\lambda[a \ _\lambda \  b]
  \ , \ [a \ _\lambda \ \partial b] = (\lambda+\partial)[a \ _\lambda \ b]
& \qquad \text{(sesquilinearity)} \\
& \hspace{-3.8cm} [b  \ _\lambda \  a] = -[a \  _{-\lambda-\partial} \ b]
& \qquad \text{(skewsymmetry)} \\
& \hspace{-.6cm} [a \ _\lambda \  [b \ _\mu \ c]]-
   [b \ _\mu  \ [a \ _\lambda \  c]] =
   [[a \ _\lambda \  b] \ _{\lambda+\mu} \ c]
& \qquad \text{(Jacobi identity)}
\end{eqnarray*}
for $a,b,c\in R$.
Here and further, $\otimes$ stands for the tensor product of vector
spaces over $\bC$.
We call \emph{rank} of a Lie conformal algebra its rank
as $\bC[\partial]$ module.
A \emph{module} $M$ over the Lie conformal algebra
$R$ is a $\bC[\partial]$-module endowed with a $\bC$-bilinear map
$R\otimes M \rightarrow \bC[\lambda]\otimes M$ denoted by $a\otimes v
\rightarrow a_\lambda v$
and called $\lambda$-action, satisfying the following axioms:
\begin{eqnarray}\label{confmod}
&& (\partial a)_\lambda v=-\lambda(a_\lambda v) \ , \
   a_\lambda (\partial v)=(\lambda+\partial)(a_\lambda v) \nonumber \\
&& a_\lambda(b_\mu v)-b_\mu(a_\lambda v)=
   [a \ _\lambda \  b]_{\lambda+\mu} v
\end{eqnarray}
for $a,b\in R$ and $v\in M$. The notions of homomorphism, ideal and
subalgebras of a Lie conformal algebra are defined in the usual way.
For a Lie conformal algebra we can define a $\bC$-bilinear product
$R\otimes R \rightarrow  R$ for any $n\in \bZ_+$, denoted by:
$a\otimes b \rightarrow a_{(n)} b$ and given by:
\begin{equation}
[a \ _\lambda \  b]=\sum_{n\in\bZ_+}\lambda^{(n)}a_{(n)}b \ ,
\end{equation}
where we are using the notation: $\lambda^{(n)}:=\frac{\lambda^n}{n!}$.
Similarly, for a conformal module over a Lie conformal algebra, we have the
corresponding notion of $n-th$ action $R\otimes M \rightarrow  M$ for any
$n\in\bZ_+$, given by: $a_\lambda v=\sum_{n\in\bZ_+}\lambda^{(n)}a_{(n)}v$.


Particularly important in physics is the \emph{Virasoro conformal
algebra}. It is defined as the free $\bC[\partial]$-module of rank 1
generated by an element $L$, with $\lambda$-bracket defined by
\begin{equation}\label{vir}
[L \ _\lambda \  L]=(\partial+2\lambda)L
\end{equation}
and extended to $\bC[\partial]\otimes L$ using sesquilinearity.
In any Lie conformal algebra $R$ we call a \emph{Virasoro element}
$L\in R$ any element satisfying
the $\lambda$-bracket relation (\ref{vir})
and such that $L_{(0)}a=\partial a \ \forall a\in R$ and $L_{(1)}$ is a
diagonalizable operator over $\bC$ in $R$.
Clearly, given a Virasoro element $L\in R$,
we construct a Virasoro conformal subalgebra in $R$ by taking
$\bC[\partial]\otimes L\subset R$.


We emphasize here the following important fact about any Virasoro element $L$
of a Lie conformal algebra $R$.
Consider the Lie algebra of polynomial vector fields in $\bC$,
\begin{equation}\label{wrepr}
W=\bigoplus_{n\geq-1}\bC L_n \ , \
\end{equation}
where $L_n=-z^{n+1}\frac{d}{dz}$, so that
$[L_m \ , \ L_n]=(m-n)L_{m+n}$.
Then we get a representation of $W$ on $R$ given by:
$$
L_n(a)=L_{(n+1)}a \ .
$$
As an immediate consequence we have a representation on $R$ of the Lie
algebra
\begin{equation}\label{sl2repr}
\operatorname{sl}_2= \text{span} \{E , F , H\} \ \subset W \ ,\
\end{equation}
where: $E=L_1 \ ,\  F=-L_{-1} \ , \ H=-2L_0$ and $H$ is a diagonalizable
operator.


In the present paper we will be concerned mainly with the most important
example of infinite rank Lie conformal algebra, the
\emph{general Lie conformal algebra} $\gc_N$ \cite[sec 2.10]{kac},
\cite{kac3}.
As a $\bC[\partial]$--module, it is defined as
$$
\gc_N=\bigoplus_{n\in\Z_+}
\bC[\partial](J^n\otimes\operatorname{Mat}_N\bC) \ .
$$
If we denote $J^n_A:=J^n\otimes A$,
the $\lambda$--bracket between two such elements is given by
$$
[J^m_A  \ _\lambda \  J^n_B]=
\sum_{j=0}^m\binom{m}{j}(\lambda+\partial)^j J^{m+n-j}_{AB}
-\sum_{j=0}^n\binom{n}{j}(-\lambda)^j J^{m+n-j}_{BA}
$$
and it is extended to the whole space $\gc_N$ using sesquilinearity.
There is a natural structure of a $\gc_N$--module on $\bC[\partial]^N$
corresponding to the action of differential operators on functions,
given by
$$
J^m_A \ _\lambda v := (\lambda+\partial)^m Av
$$
for any $v\in\bC^N$, and extended to $\bC[\partial]^N$ using
axioms (\ref{confmod}).


It is convenient to identify $\gc_N$ with the space of
$N\times N$ matrices with entries being polynomials in $\partial$ and $x$
$$
\gc_N \xrightarrow{\sim} \operatorname{Mat}_N\bC[\partial,x]
$$
via the so called \emph{symbol map}: $\partial^k J^m_A \mapsto
\partial^k x^m A$.
With this notation the $\lambda$--bracket structure of $\gc_N$ becomes
\begin{eqnarray}\label{xbracket}
[A(\partial,x) \ _\lambda \  B(\partial,x)]
&=&A(-\lambda,\lambda+\partial+x)\cdot B(\lambda+\partial,x) \\
&&-B(\lambda+\partial,-\lambda+x)\cdot A(-\lambda,x) \ , \nonumber
\end{eqnarray}
where $A(\partial,x)$ and $B(\partial,x)\in \operatorname{Mat}_N
\bC[\partial,x]$ and $\cdot$ denotes the product of $N\times N$ matrices.
The $\lambda$--action of $\gc_N$ on $\bC[\partial]^N$
becomes
\begin{equation}\label{xaction}
A(\partial,x)_\lambda v(\partial)=A(-\lambda,\lambda+\partial)\cdot
v(\lambda+\partial) \ ,
\end{equation}
where $A(\partial,x)\in \operatorname{Mat}_N\bC[\partial,x]$,
$v(\partial)\in \bC[\partial]^N$ and now $\cdot$ denotes the action
of $N\times N$ matrices on $N$-vectors.
Finally, we will also use the following change of variables
\begin{equation}\label{yvar}
(\partial,x)\mapsto (\partial,y=2x+\partial)
\end{equation}
so that the $\lambda$--bracket (\ref{xbracket}) takes
the more symmetric form
\begin{eqnarray*}
[A(\partial,y) \ _\lambda \  B(\partial,y)]
&=&A(-\lambda,y+\lambda+\partial)\cdot
B(\lambda+\partial,y+\lambda) \\
&&-B(\lambda+\partial,y-\lambda)\cdot
A(-\lambda,y-\lambda-\partial) \ . \nonumber
\end{eqnarray*}
The Lie conformal algebra $\gc_N$ plays the same role in the theory of
Lie conformal algebras as $\operatorname{gl}_N$ does in the theory of
Lie algebras: any module $M=\bC[\partial]^N$ over a
Lie conformal algebra $R$ is obtained via a homomorphism
$R\rightarrow\gc_N$, \cite{kac}, \cite{kac3}.


It is an easy exercise to prove that all elements of the form
$L=(x+\alpha\partial)\id$ with $\alpha\in\bC$ are Virasoro elements
of $\gc_N$.
Here and further $\id$ stands for the identity matrix.
In terms of the $y$ variable introduced in (\ref{yvar}), it becomes
\begin{equation}\label{virel}
L=(x+\alpha\partial)\id=\frac{1}{2}(y-\sigma\partial)\id \
, \ \text{ where } \sigma=1-2\alpha \ .
\end{equation}
These are all Virasoro elements in $\gc_1$, but already for $N=2$
the complete list is quite complicated \cite{kac2}, and the answer
is unknown for $N>2$.
Note also that all Virasoro subalgebras of $\gc_1$ are conjugate to those
generated by Virasoro elements.


One knows the classification of finite Lie conformal subalgebras $R$ of
$\gc_N$ that act irreducibly on $\bC[\partial]^N$: $R$ is conjugate
to a subalgebra of the finite subalgebra of $\gc_N$
consisting of the elements $a(\partial)+\alpha x\id$, where
$a(\partial)\in\operatorname{Mat}_N\C[\partial]$, $\alpha\in\C$
\cite[Theorem 8.6]{kac3}.


The problem of classifying all infinite rank subalgebras of $\gc_N$
that act irreducibly on $\bC[\partial]^N$ is still open.
In the present paper we classify all such subalgebras
which are normalized with respect to
a Virasoro element $L$ of kind (\ref{virel}),
in the sense defined below.
In particular, we classify all irreducible subalgebras of $\gc_N$
that contain a Virasoro element (\ref{virel})
(they are the most interesting $\gc_N$-subalgebras from the
point of view of physics).
The analogous, though easier, problem of  classification of
infinite rank
subalgebras of the associative conformal algebra $Cend_N$ has been
solved in \cite{liberati} for $N=1$ (but the problem is still open for
$N>1$).
In that paper a complete list of infinite rank subalgebras of $\gc_N$
is also conjectured and our result gives a partial confirmation
of that conjecture.


In order to classify subalgebras of the general conformal algebra $\gc_N$
containing a given Virasoro element $L$, we will use the so called
quasi--primary elements.
An element $a\in \gc_N$ is called a \emph{primary element}
(with respect to the Virasoro element $L$) of conformal weight
$\Delta$ if:
$$
[L \ _\lambda \  a]=(\partial + \Delta\lambda)a \ .
$$
This has a nice interpretation in terms of the $W$-module structure of $\gc_N$
defined by (\ref{wrepr}). A primary element $a\in \gc_N$ is any element which
is annihilated by the subalgebra
$\n_+=\bigoplus_{n\geq +1}\bC L_n$ and is an eigenvector of $L_0$, the
eigenvalue being the conformal weight
$\Delta$. Clearly $L$ itself is a primary element of conformal weight
$\Delta=2$.
More generally one defines a \emph{quasi--primary element} to be an element
$a\in \gc_N$ such that:
$$
[L \ _\lambda \  a]=\partial a + \lambda \Delta a + O(\lambda^3)
$$
In other words $a$ is a highest weight vector of the Lie
algebra $\operatorname{sl}_2 =$ span$\{E,F,H\}$ defined above, with highest
weight $-2\Delta$.


Notice that for the particular choice of $L$ as in (\ref{virel})
one gets immediately
$$
L_{(1)}a=(n+1)a
$$
as soon as  $a\in\operatorname{Mat}_N\bC[\partial,x]
\simeq\operatorname{Mat}_N\bC[\partial,y]$
is a homogeneous polynomial in $\partial$, $x$ (or, equivalently,
in $\partial$, $y$) of degree $n$.
Hence $L_0=L_{(1)}$ is diagonal in the basis of monomials
$\partial^kx^ne_{ij}$ (or equivalently $\partial^ky^ne_{ij}$),
and the eigenvalues are all positive integers.
In other words, the Cartan element $H\in\operatorname{sl}_2$
is diagonalizable on $\gc_N$ with
negative even integer eigenvalues.
We can apply the following well known result in representation
theory of $\operatorname{sl}_2$:
\begin{lemma}
Suppose $V$ is an $\operatorname{sl}_2$--module such that $H$ is
diagonalizable with {\upshape spec}$(H)<$ {\upshape const}
and {\upshape spec}$(H)\cap\Z_+=\emptyset$. Then $V$
decomposes in a direct sum of irreducible Verma modules.
\end{lemma}
\begin{corollary}\label{decomp}
For $L$ defined by (\ref{virel}), $\gc_N$ is a direct sum of irreducible
Verma modules over $\operatorname{sl}_2$.
\end{corollary}
Any element in $\gc_N$ is therefore obtained starting with
quasi--primary elements
and applying the ``lowering operator'' $L_{-1}=\partial$.
Conversely, starting with any element $a\in \gc_N$ we get a quasi--primary
element by applying a power of the ``raising operator'' $L_1$.


Fix a Virasoro element $L$ of the form (\ref{virel}).
The \emph{reduced space} $\hat{\gc}_N$ (with respect to
$L$) is the $\bC$--span
of all quasi--primary elements. It follows by the previous remarks
that any element $a\in \gc_N$ can be decomposed uniquely in the
following way:
\begin{equation}\label{dec}
a=\sum_{i\geq 0}\partial^i a^i \ ,
\end{equation}
where $a^i\in\hgc_N$ and only finitely many terms are non zero.
This defines a projection map $\pi \ : \ R\rightarrow\hat{\gc}_N$
by taking $\pi(a)=a^0$.


We now want to define in the reduced space $\hgc_N$ a structure induced by
the conformal algebra structure of $\gc_N$.
Following the idea introduced in \cite{yamamoto}, for any $n\geq 0$ we define
a product $\hgc_N\otimes\hgc_N\rightarrow\hgc_N$, denoted by
$a\otimes b \rightarrow a_{<n>}b$, in the following way: first we take the
$n$-th product $a_{(n)}b$ in $\gc_N$ and then we project on $\hgc_N$.
Namely, if $a_{(n)}b$ decomposes as in (\ref{dec}):
$a_{(n)}b=\sum_{i\geq 0}\partial^i (a_{(n)}b)^i$, we define
$$
a_{<n>}b=(a_{(n)}b)^0 \ .
$$
We will denote by $[a_{<\lambda>}b]$ the corresponding $\lambda$-bracket
in the reduced space:
\begin{equation}\label{redbra}
[a_{<\lambda>}b]=\sum_{n\geq 0} \lambda^{(n)} a_{<n>}b
= \pi\left([a \ _\lambda \ b]\right) \ .
\end{equation}
\begin{remark}\label{yamskwesym}
The skew symmetry of the $\lambda$--bracket implies that
$$
a_{<n>}b=(-1)^{n+1}b_{<n>}a \ .
$$
One can also write down the identity corresponding to the Jacobi identity
\cite{yamamoto}, but we will not need it.
\end{remark}


A conformal subalgebra $R\subset\gc_N$ is called a
\emph{normalized} subalgebra (with respect $L$),
if $L_{(n)}R\subset R$ for $n=0,1,2$, or, equivalently,
$R$ is a submodule of the Lie algebra
$\operatorname{sl}_2\subset W$, defined in (\ref{sl2repr}).


We shall establish a natural correspondence between normalized
subalgebras $R\subset\gc_N$ (with respect to $L$)
and subalgebras $\hat{R}$ of the space $\hgc_N$
(reduced with respect to the same $L$).
\begin{lemma}\label{lemma1}
Let $R\subset\gc_N$ be any normalized subalgebra. \\
{\bfseries (i)} For $a\in R$, all the coefficients $a^i$ of the
decomposition (\ref{dec}) are in $R\cap \hgc_N$.
In particular $\pi(R)=R\cap \hgc_N$. \\
{\bfseries (ii)} Moreover, the projection $\pi(R)$ is a subalgebra of
the reduced space $\hgc_N$, namely it is closed with respect to
all $n$-th products defined in $\hgc_N$.
\end{lemma}
\begin{proof}
The first statement follows from Corollary \ref{decomp}.
For (ii), let $a$, $b\in\pi(R)=R\cap\hgc_N$.
By definition $a_{<n>}b=(a_{(n)}b)^0$, and this is again
in $R\cap\hgc_N$ by part (i).
\end{proof}
\begin{lemma}\label{lemma2}
Let $\hat{R}\subset\hgc_N$ be a subalgebra of the reduced space. \\
{\bfseries (i)} Suppose $a$, $b\in\hat{R}$  are quasi--primary elements
of conformal weights $\Delta(a)$ and $\Delta(b)$ respectively.
For $n\in\Z_+$ consider the elements $(a_{(n)}b)^i\in\hgc_N$ of
the decomposition (\ref{dec}). One has
$$
(a_{(n)}b)^i=C a_{<n+i>}b \ ,
$$
where $C\in\C$ is some constant depending on $\Delta(a)$, $\Delta(b)$,
$n$ and $i$.
In particular $(a_{(n)}b)^i\in\hat{R}$ for all $i\geq0$. \\
{\bfseries (ii)} $\C[\partial]\hat{R}$ is a
normalized subalgebra of $\gc_N$.
\end{lemma}
\begin{proof}
(i) is obtained by applying recursively $L_{(2)}$ to both sides
of the decomposition $a_{(n)}b=\sum_{i\geq0}\partial^i(a_{(n)}b)^i$.
We omit this calculation, which can be found in \cite{yamamoto}.
By (i) we then have that $a_{(n)}b\in\C[\partial]\hat{R}$ for all $n\in\Z_+$.
But then $\hat{R}_{(n)}\hat{R}\subset\C[\partial]\hat{R}$, so that,
by sesquilinearity, $C[\partial]\hat{R}$ is a conformal subalgebra of $\gc_N$,
thus proving (ii).
\end{proof}
As immediate consequence of Lemma \ref{lemma1} and Lemma \ref{lemma2}
we get the following result:
\begin{corollary}\label{baslem}
Given a Virasoro element $L$ of the form (\ref{virel}),
the maps $\phi : \ R\longrightarrow \pi(R)$
and $\phi^{-1} : \ \hat{R}\longrightarrow \C[\partial]\hat{R}$
give a bijective correspondence between normalized subalgebras
$R\subset\gc_N$ and subalgebras $\hat{R}$ of the reduced space $\hgc_N$.
\end{corollary}


In the following sections we will fix a Virasoro element
$L=(x+\alpha\partial)\id =\frac{1}{2}(y-\sigma\partial)\id$
in the general conformal algebra $\gc_N$
and we will study the algebra structure of the corresponding reduced
space $\hgc_N$.
Remarkably, we find that the quasi--primary elements turn out to
be the well-known Jacobi polynomials (in homogeneous form).
We will use this fact and Corollary \ref{baslem} in order to classify
the subalgebras of $\gc_N$ which are normalized with respect to $L$ and
which act irreducibly on $\C[\partial]^N$.
To simplify notation, we will consider first the case of $\gc_1$.
The generalization to $\gc_N$, for any positive integer $N$, will be
discussed in Section 4.


\ssection{Algebra structure of the reduced space $\hgc_1$}

\ssubsection{Basis for $\hgc_1$}


\noindent
We consider the general conformal algebra $\gc_1=\bC[\partial,x]$
and we fix a Virasoro element $L=x+\alpha\partial$.
Our first task will be to classify all quasi--primary elements in $\gc_1$,
i.e. all elements $A(\partial,x)\in\bC[\partial,x]$
such that $L_{(2)}A(\partial,x)=0$.


As a vector space over $\bC$, $\gc_1$ has a basis consisting of
the elements $\partial^k x^n$.
By (\ref{xbracket}) we have:
$$
[L \ _\lambda \  \partial^k x^n]
=\left((1-\alpha)\lambda+\partial+x\right)
    (\lambda+\partial)^k x^n
   - (-\alpha\lambda+x)(\lambda+\partial)^k(-\lambda+x)^n \ .
$$
After a straightforward computation we get the (2)-nd product:
\begin{eqnarray}\label{2ndpr}
&& L_{(2)}(\partial^k x^n) =  \left.\frac{d^2}{d\lambda^2}
   [L \ _\lambda \  (\partial^k x^n)]\right|_{\lambda=0} \\
&& = (k(k+1)+2kn)\partial^{k-1}x^n-(n(n-1)+2n\alpha)\partial^kx^{n-1}  \ .
\nonumber
\end{eqnarray}
By an induction argument, we then get a basis
$\{Q^{(\sigma)}_n(\partial,x) \ , \ n\in\Z_+\}$
for the space of quasi--primary elements by taking:
$$
Q^{(\sigma)}_n(\partial,x)=\sum_{k=0}^n c_{n,k}\partial^k x^{n-k} \ ,
$$
where the coefficients $c_{n,k}$ are such that the condition of
quasi--primarity is satisfied:
$$
0=L_{(2)}Q^{(\sigma)}_n(\partial,x)=\sum_{k=0}^n c_{n,k}
L_{(2)}(\partial^k x^{n-k}) \ .
$$
This condition becomes, after using (\ref{2ndpr}) and some simple algebraic
manipulations, the following recursive relation on the coefficients $c_{n,k}$
$$
c_{n,k}k(2n-k+1)=c_{n,k-1}(n-k+1)(n-k+2\alpha) \ ,
\ \forall n\geq0 \ , \ k=1,\dots,n \ .
$$
The choice of the leading coefficient $c_{n,0}$ is arbitrary. If we fix
$c_{n,0}=1$ we get at once all other coefficients
$$
c_{n,k}
   =\frac{\binom{2n-k}{n}\binom{n+2\alpha-1}{k}}{\binom{2n}{n}} \ .
$$
Thus we have proved the following
\begin{theorem}\label{reducedspace}
The space $\hgc_1\subset\gc_1$, reduced with respect to the
Virasoro element $L=x+\alpha\partial$, is a vector space over $\bC$
with basis
\begin{equation}\label{qn}
Q_n^{(\sigma)}(\partial,x) = \frac{1}{\binom{2n}{n}}
\sum_{k=0}^n \binom{2n-k}{n}\binom{n-\sigma}{k} \partial^k x^{n-k} \ ,
\end{equation}
where $\ n\in\Z_+$ and $\sigma=1-2\alpha$.
We can write these polynomials in terms of the $y$ variable
defined in (\ref{yvar}), so that a basis of the reduced space $\hgc_1$
is given by the following polynomials in $\partial$ and $y$
\begin{equation}\label{rn}
R_n^{(\sigma)}(\partial,y)=
Q_n^{(\sigma)}\left(\partial,\frac{1}{2}(y-\partial)\right) \ .
\end{equation}
\end{theorem}
The first two basis elements are $Q_0^{(\sigma)}(\partial,x)=1$ and
$Q_1^{(\sigma)}(\partial,x)=x+\alpha\partial=L$.
The basis (\ref{qn}) for the space of quasi--primary elements was first found in
 \cite{bakalov} in the particular situation $\sigma=0$,
i.e. for the particular choice of the Virasoro element
$L=x+\frac{1}{2}\partial=\frac{1}{2}y$.

\ssubsection{Relation with Jacobi polynomials}


\noindent
It is an interesting fact that the polynomials $Q_n^{(\sigma)}(\partial,x)$
defined by (\ref{qn}) are closely related to the Jacobi polynomials.
Let us briefly recall the definition and the main properties of
Jacobi polynomials.
For a more exhaustive discussion on special functions and orthogonal systems,
see, for example, \cite{jacobi}.
Given two parameters $\alpha$ and $\beta$, the \emph{Jacobi polynomial}
$P^{(\alpha,\beta)}_n$ of degree $n$ can be defined by:
\begin{equation}\label{jac}
P^{(\alpha,\beta)}_n(y):=\binom{\alpha+n}{n}
F\left(-n,n+\alpha+\beta+1;\alpha+1;\frac{1-y}{2}\right) \ ,
\end{equation}
where the hypergeometric function $F(a,b;c;x)$ is given by
\begin{equation}
F(a,b;c;x)
= \sum_{n\geq0}
\frac{\binom{a+n-1}{n}\binom{b+n-1}{n}}{\binom{c+n-1}{n}}x^n \ .
\end{equation}

The generating function for Jacobi polynomials is given by
\begin{equation}\label{jacgen}
F^{(\alpha,\beta)}(y,w):=\sum_{n\geq0}P^{(\alpha,\beta)}_n(y)w^n
=2^{\alpha+\beta} R^{-1}
[1-w+R]^{-\alpha} [1+w+R]^{-\beta}
\end{equation}
where
$$
R=\sqrt{1-2yw+w^2} \ .
$$
Another way to define Jacobi polynomials is as solutions of second order
differential equations. More precisely, the Jacobi polynomial
$P^{(\alpha,\beta)}_n(y)$ is the unique polynomial of degree $n$
with leading coefficient $\frac{1}{2^n}\binom{2n+\alpha+\beta}{n}$
which is a solution of the differential equation
\begin{equation}\label{jacode}
(1-y^2)u^{\prime\prime}+[\beta-\alpha-(\alpha+\beta+2)y]u^\prime
+n(n+\alpha+\beta+1)u = 0 \ .
\end{equation}
In the following, we will also need the symmetry relation between
Jacobi polynomials
\begin{equation}\label{jacsym}
P^{(\alpha,\beta)}_n(y)=(-1)^nP^{(\beta,\alpha)}_n(-y) \ .
\end{equation}


Now we want to relate the quasi--primary elements discussed in
the previous section to Jacobi polynomials.
Starting by the definition (\ref{qn}), it is an easy matter to check
the following relation between the polynomials $Q_n^{(\sigma)}(\partial,x)$
and the hypergeometric function $F(a,b;c;x)$
$$
\binom{2n}{n}Q_n^{(\sigma)}(-1,x)=(-1)^n\binom{n-\sigma}{n}
F(-n,n+1;-\sigma+1;x) \ .
$$
But then it follows by the definition (\ref{jac}) of Jacobi polynomials
and by the fact that $Q_n^{(\sigma)}(\partial,x)$ is a homogeneous polynomial
in $\partial$ and $x$, that the following interesting relation holds
\begin{equation}\label{qnpn}
\binom{2n}{n}Q_n^{(\sigma)}(\partial,x)=\partial^n
P^{(-\sigma,\sigma)}_n\left(\frac{2}{\partial}(x+\frac{1}{2}\partial)\right)
\ .
\end{equation}
This relation takes a nicer form in terms of the $y$ variable defined
in (\ref{yvar}) and the polynomials $R_n^{(\sigma)}(\partial,y)$
defined in (\ref{rn}):
\begin{equation}\label{r-p}
\binom{2n}{n}R_n^{(\sigma)}(\partial,y)=\partial^n
P^{(-\sigma,\sigma)}_n\left(\frac{y}{\partial}\right) \ .
\end{equation}


Notice how relation (\ref{r-p}) can also be derived by using the
fact that Jacobi polynomials are solutions of the differential
equation (\ref{jacode}).
The $\lambda$--bracket of
$L=\frac{1}{2}(y-\sigma\partial)$ with a generic element
$q(\partial,y)\in\gc_N$ is:
\begin{eqnarray}\label{virlq}
[L \ _\lambda \ q(\partial,y)]&=&
\frac{1}{2}((1+\sigma)\lambda+\partial+y)q(\lambda+\partial,\lambda+y) \\
&+& \frac{1}{2}((1-\sigma)\lambda+\partial-y)q(\lambda+\partial,-\lambda+y)
\ . \nonumber
\end{eqnarray}
Looking at this expression as a polynomial in $\lambda$, we clearly have
that the constant term is $L_{(0)}q(\partial,y)=\partial q(\partial,y)$.
Now let's consider the coefficient of $\lambda$. We get
$$
L_{(1)}q(\partial,y)=
\left.\frac{d}{d\lambda}[L \ _\lambda \ q(\partial,y)]\right|_{\lambda=0}
= (1+\partial D_1+y D_2)q(\partial,y) \ ,
$$
where $D_1$ (resp. $D_2$) denotes derivative with respect to the first
(resp. second) variable.
In particular, if we assume that $q(\partial,y)$ is a homogeneous polynomial
in $\partial$ and $y$ of degree $n$, it is an eigenvector of $L_{(1)}$
with eigenvalue $n+1$.
In this case $q(\partial,y)$ satisfies the identity
\begin{equation}\label{d12}
(\partial D_1+y D_2 -n)q(\partial,y)=0 \ .
\end{equation}
By imposing that the coefficient of $\lambda^2$ in (\ref{virlq}) be equal
to zero, we get a necessary and sufficient condition for $q(\partial,y)$ to be
a quasi--primary element:
\begin{eqnarray*}
&&L_{(2)}q(\partial,y)=
\left.\left(\frac{d}{d\lambda}\right)^2
[L \ _\lambda \ q(\partial,y)]\right|_{\lambda=0} \\
&&\ = [\partial D_1^2+\partial D_2^2 +2yD_1D_2 +2D_1+2\sigma D_2]
q(\partial,y) =0 \ .
\end{eqnarray*}
But then just multiplying by $\partial$ and using relation (\ref{d12})
we get the differential equation
$$
[(\partial^2-y^2)D_2^2 + 2(\sigma\partial-y)D_2 +n(n+1)]q(\partial,y)=0 \ ,
$$
which, after putting $\partial=1$, is the same as equation (\ref{jacode})
in the particular case $\beta=-\alpha=\sigma$.
Considering that $q(\partial,y)$ is homogeneous
in its variables, this implies, apart from a constant factor,
that the quasi--primary element $R^{(\sigma)}_n(\partial,y)$ is related
to the Jacobi polynomial $P^{(-\sigma,\sigma)}_n(y)$ by (\ref{r-p}).


We point out the following interesting symmetry relation
satisfied by the polynomials $R_n^{(\sigma)}(\partial,y)$, which is an
immediate consequence of (\ref{jacsym}) for Jacobi polynomials and (\ref{r-p}):
$$
R_n^{(\sigma)}(\partial,y)=R_n^{(-\sigma)}(-\partial,y) \ .
$$
In terms of the variable $x$ and the polynomials $Q_n^{(\sigma)}(\partial,x)$,
this same relation becomes
\begin{equation}\label{sym}
Q_n^{(\sigma)}(\partial,x)=Q_n^{(-\sigma)}(-\partial,\partial+x) \ .
\end{equation}


We also find the generating function for the polynomials
$R_n^{(\sigma)}(\partial,y)$ by using the expression (\ref{jacgen})
of the generating function for Jacobi polynomials. The result
is the following
\begin{eqnarray}\label{ygener}
&& R^{(\sigma)}(\partial,y,z):=\sum_{n\geq0}
\binom{2n}{n}R_n^{(\sigma)}(\partial,y)z^n \\
&& = \frac{1}{\sqrt{1-2yz+\partial^2z^2}}
\left[\frac{1-\partial z+\sqrt{1-2yz+\partial^2z^2}}
{1+\partial z+\sqrt{1-2yz+\partial^2z^2}}\right]^\sigma \ . \nonumber
\end{eqnarray}
\ssubsection{Products in $\hgc_1$}

\noindent
In order to find the algebra structure of the reduced space
we need to study the $\lambda$-bracket among elements in $\hgc_1$.
It is sufficient to consider the basis elements
$Q_{n}^{(\sigma)}(\partial,x)$ defined in (\ref{qn}).
By (\ref{xbracket}) we have
\begin{eqnarray}\label{lambda}
[Q_{m}^{(\sigma)}(\partial,x)\ _\lambda \ Q_{n}^{(\sigma)}(\partial,x)]
&=& Q_m^{(\sigma)}(-\lambda,\lambda+\partial+x)
   Q_n^{(\sigma)}(\lambda+\partial,x) \nonumber\\
&-& Q_m^{(\sigma)}(-\lambda,x)Q_n^{(\sigma)}(\lambda+\partial,-\lambda+x) \ .
\end{eqnarray}
Notice that we are not interested in the entire expression of this
$\lambda$-bracket, but only in its projection on the reduced
space $\hgc_1$. In order to get it we can use the following:
\begin{lemma}\label{trivlemma}
The projection $\pi \ : \ \gc_1\rightarrow \hgc_1$ is obtained by first
putting $\partial=0$ and then replacing $x^n$ by
$Q_n^{(\sigma)}(\partial,x)$, namely
\begin{equation}\label{proj}
\left.\pi(A(\partial,x))=A(0,x)
\right|_{x^n\rightarrow Q_n^{(\sigma)}(\partial,x)} \ .
\end{equation}
\end{lemma}
\begin{proof}
Suppose $A(\partial,x)$ decomposes as in (\ref{dec})
$$
A(\partial,x) = \sum_{k\geq0} \partial^k
\left(\sum_{n\geq0} c_n^{(k)} Q_n^{(\sigma)}(\partial,x)\right) \ .
$$
Then its projection is
$$
\pi\left(A(\partial,x)\right) = \sum_{n\geq0}
c_{n}^{(0)} Q_n^{(\sigma)}(\partial,x) \ .
$$
On the other hand, by putting $\partial=0$ in the expression of
$A(\partial,x)$ we get
$$
A(0,x) =
\sum_{n\geq0} c_{n}^{(0)} x^n
$$
since $Q_n^{(\sigma)}(\partial,x)$ are defined such that
$Q_n^{(\sigma)}(0,x)=x^n$.
By comparing the last two equalities we immediately get (\ref{proj}).
\end{proof}
If we put $\partial=0$ in the right hand side of (\ref{lambda}) we get, after
using the symmetry relation (\ref{sym}):
$$
 Q_m^{(-\sigma)}(\lambda,x)Q_n^{(\sigma)}(\lambda,x)
 - Q_m^{(\sigma)}(-\lambda,x)Q_n^{(-\sigma)}(-\lambda,x) \ .
$$
Therefore by Lemma \ref{trivlemma} we just need to replace $x^n$ with
$Q_n^{(\sigma)}(\partial,x)$ in order to get the $\lambda$-bracket structure of
the reduced space $\hgc_1$:
\begin{eqnarray}\label{mln}
&&[Q_{m}^{(\sigma)}(\partial,x) _{<\lambda>} Q_{n}^{(\sigma)}(\partial,x)] \\
&&\qquad = Q_m^{(-\sigma)}(\lambda,x)Q_n^{(\sigma)}(\lambda,x)
- \left.Q_m^{(\sigma)}(-\lambda,x)Q_n^{(-\sigma)}(-\lambda,x)
   \right|_{x^n\rightarrow Q_n^{(\sigma)}(\partial,x)} \nonumber
\end{eqnarray}
By (\ref{redbra}) the $k$-th product is the coefficient of
$\lambda^{(k)}$ in this expression.
This defines completely the algebra structure of the reduced space $\hgc_1$.


It is convenient to introduce the following simplified notation
for the reduced space $\hgc_1$.
We define the isomorphism
$$
\hgc_1\xrightarrow{\sim}\bC[X]
$$
obtained by identifying $Q_n^{(\sigma)}(\partial,x)= X^n$
and by taking on $\bC[X]$ the induced algebra structure.
The expression (\ref{mln}) for the $\lambda$--bracket in $\hgc_1$
thus becomes
\begin{equation}\label{mln2}
[X^m \ _{<\lambda>}\ X^n]=
Q_m^{(-\sigma)}(\lambda,X)Q_n^{(\sigma)}(\lambda,X)
- Q_m^{(\sigma)}(-\lambda,X)Q_n^{(-\sigma)}(-\lambda,X)
\end{equation}


We now want to find an explicit expression for all $k$-th products. For this,
notice that in the right hand side of (\ref{mln2}) the two terms are
obtained one from each other by exchanging
$$
\sigma\rightarrow-\sigma \ , \ \lambda\rightarrow-\lambda \ .
$$
Therefore it is sufficient to analyze only the first term. A straightforward
computation gives:
$$
Q_m^{(-\sigma)}(\lambda,X)Q_n^{(\sigma)}(\lambda,X)
= \sum_{k=0}^{m+n} \frac{\lambda^k}{k!}d^{(\sigma)}_{m,n,k} X^{m+n-k} \ ,
$$
where:
\begin{equation}\label{dmnk}
d^{(\sigma)}_{m,n,k}=
\frac{k!}{\binom{2m}{m}\binom{2n}{n}}
\sum_{i,j:\atop{0\leq i\leq m\atop{0\leq j\leq n\atop{i+j=k}}}}
   \binom{2m-i}{m}\binom{m+\sigma}{i}\binom{2n-j}{n}\binom{n-\sigma}{j} \ .
\end{equation}
Using this result we get immediately by Lemma \ref{trivlemma}
the following
\begin{theorem}\label{allprod}
The $k$-th product of basis elements in the reduced space
$\hgc_1\simeq\bC[X]$ is given by
\begin{equation}\label{kth}
X^m _{<k>} X^n
= [d^{(\sigma)}_{m,n,k} +(-1)^{k+1} d^{(-\sigma)}_{m,n,k}] X^{m+n-k} \ ,
\end{equation}
where the coefficients $d^{(\sigma)}_{m,n,k}$ are given by (\ref{dmnk}).
\end{theorem}
\begin{remark}\label{remr0}
Notice that coefficients $d^{(\sigma)}_{m,n,k}$ satisfy the symmetry
relation (cf. Remark \ref{yamskwesym}):
$d^{(\sigma)}_{m,n,k}=d^{(-\sigma)}_{n,m,k}$.
\end{remark}


One can compute explicitly products for low values of $k$.
We can assume, by Remark \ref{remr0}, $m\leq n$.
For $k=0,\dots,4$ one gets:
\begin{eqnarray}\label{keq3}
X^m _{<0>} X^n
&=& 0 \qquad \forall m,n \geq0 \ , \nonumber\\
X^m _{<1>} X^n
&=& (m+n)X^{m+n-1} \qquad \forall m,n\geq0\
\text{ such that }  m+n\geq1 \ ,\nonumber\\
X^m _{<2>} X^n
&=& 0 \qquad \forall m,n \geq1 \ , \\
X^0 _{<2>} X^n
&=& -\sigma(n-1)X^{n-2} \qquad \forall n\geq2 \ , \nonumber\\
X^m _{<3>} X^n
&=& \frac{(m+n-1)(m+n-2)}{2(2m-1)(2n-1)} [2m^2n+2n^2m-m^2-n^2-5mn \nonumber\\
&& +2m+2n-3\sigma^2] X^{m+n-3} \quad \forall m,n\geq0\
\text{ such that } m+n\geq3 \ , \nonumber\\
X^m _{<4>} X^n
&=& 0 \qquad \forall m,n \geq2 \ , \nonumber\\
X^1 _{<4>} X^n
&=& -\sigma(1-\sigma^2)(n-2) X^{n-3} \qquad \forall n\geq3 \ , \nonumber\\
X^0 _{<4>} X^n
&=& -\sigma\frac{(n-2)(n-3)}{2n-1}
(n^2-3n+\sigma^2+1) X^{n-4} \qquad \forall n\geq4 \ . \nonumber
\end{eqnarray}
It is also possible to compute explicitly products for values of $k$ which
are close to $m+n$. For $k=m+n$ one gets:
$$
X^m _{<m+n>} X^n = \frac{(m+n)!}{\binom{2m}{m}\binom{2n}{n}}
\left[\binom{m+\sigma}{m}\binom{n-\sigma}{n}-(-1)^{m+n}
\binom{m-\sigma}{m}\binom{n+\sigma}{n}\right] X^0 \ .
$$
In particular, for $n=m,\ m+1,\ m+2$ one has:
\begin{eqnarray}\label{keq2m2}
X^m _{<2m>} X^m
&=& 0 \ , \nonumber\\
X^m _{<2m+1>} X^{m+1}
&=& \frac{m+1}{\binom{2m}{m}} \prod_{k=1}^m (k^2-\sigma^2) \ X^0 \ , \\
X^m _{<2m+2>} X^{m+2}
&=& -\sigma\frac{m+1}{2\binom{2m}{m}}\prod_{k=1}^m (k^2-\sigma^2)
\ X^0 \ .\nonumber
\end{eqnarray}
We will also need the expression for the $k$--th product in
the particular case of $n=m\geq1$ and $k=2n-1$.
In this situation we get:
\begin{equation}\label{keq2m1}
X^m _{<2m-1>} X^{m}=
\frac{2(m+1)}{\binom{2m}{m}} \prod_{k=2}^m (k^2-\sigma^2) \ X^1 \ .
\end{equation}


In the next section we will use Theorem \ref{allprod} in order to
classify subalgebras of $\hgc_1$ and $\gc_1$.


\ssection{Normalized subalgebras of $\gc_1$}


\noindent
We want to classify all subalgebras of $\gc_1$ which are normalized
with respect to a given Virasoro element $L=x+\alpha\partial$ and act
irreducibly on $\bC[\partial]$.
Finite subalgebras have been classified in \cite{kac3}, so that
it will be sufficient to consider only infinite rank subalgebras.


The main technique consists in looking at the space $\hgc_1\simeq\bC[X]$,
reduced with respect to the same Virasoro element $L$.
By Corollary \ref{baslem} every normalized subalgebra
$R\subset\gc_1$ corresponds
(via the projection map $\pi$) to a subalgebra $\hat{R}\subset\hgc_1$
of the corresponding reduced space.
In particular $R\subset\gc_1$ is of infinite rank as $\C[\partial]$--module
if and only if the corresponding subalgebra $\hat{R}\subset\hgc_1$ is
infinite--dimensional over $\C$.
Our goal is therefore equivalent to classifying all infinite--dimensional
subalgebras of the reduced space $\hgc_1$.


The following simple fact is very useful in studying
subalgebras of $\hgc_1$.
\begin{lemma}\label{hrdec}
Any subalgebra $\hat{R}$ of the reduced space $\hgc_1$
decomposes as
\begin{equation}\label{decbasis}
\hat{R}=\bigoplus_{n\in I} \bC X^n \ ,
\end{equation}
for some index set $I\subset\Z_+$.
\end{lemma}
\begin{proof}
Any normalized subalgebra $R\subset\gc_1$ is a module
of the Lie algebra $\operatorname{sl}_2$ defined in (\ref{sl2repr}).
It thus follows by Corollary \ref{decomp} and general arguments
in representation theory that $R$ decomposes as direct sum of
weight subspaces:
$R=\bigoplus_{n\in\Z_+}(R\cap\gc_1[n+1])$, where $\gc_1[n+1]$ denotes the
eigenspace of $L_0$ with weight $n+1$, namely the space of homogeneous
polynomials in $\partial$ and $x$ of degree $n$.
Therefore, as immediate consequence of Corollary \ref{baslem},
any subalgebra $\hat{R}$ of the reduced space $\hgc_1$
decomposes accordingly, namely as in (\ref{decbasis}).
\end{proof}
According to this Lemma, we describe completely a subalgebra
$\hat{R}\subset\hgc_1$ once we specify the collection
$\{X^n , \ n\in I\}$ of basis elements which are in $\hat{R}$.


In order to classify all infinite--dimensional subalgebras
$\hat{R}\subset\hgc_1$ we need to notice the following facts.
\begin{lemma}\label{remark1}
If $X^0\in\hat{R}$ then $\hat{R}=\hgc_1$.
\end{lemma}
\begin{proof}
For this, just notice that $X^0 _{<1>} X^{n+1} = (n+1)X^n\neq0$
for any $n\in\Z_+$ so that $X^{n+1}\in\hat{R}$ implies $X^n\in\hat{R}$.
The claim follows by the fact that, since $\hat{R}$ is infinite
dimensional, there will be some element $X^n\in\hat{R}$ for
arbitrarily large $n$.
\end{proof}
\begin{lemma}\label{remark2}
For $\sigma\notin\Z$ there are no proper infinite--dimensional
subalgebras of $\hgc_1$.
\end{lemma}
\begin{proof}
Let $\hat{R}\subset\hgc_1$ be an infinite--dimensional subalgebra,
and suppose $X^m\in\hat{R}$ for some $m>>0$.
By (\ref{keq2m1}) we get that $X^m _{<2m-1>} X^{m}\propto X^1$
with non zero coefficient, so that $X^1\in\hat{R}$.
By (\ref{keq3}) we get that $X^1 _{<3>} X^m \propto X^{m-2}$
with coefficient $\frac{m(m-1)}{2(2m-1)} [m^2-m+1-3\sigma^2]$,
which will be nonzero for $m$ large enough.
Therefore $X^{m-2}\in\hat{R}$.
Finally by (\ref{keq2m1}) we get that
$X^m _{<2m+2>} X^{m+2}\propto X^0$ with nonzero coefficient.
But then $X^0\in\hat{R}$ so that, by Lemma \ref{remark2},
$\hat{R}=\hgc_1$.
\end{proof}
In the following we assume that $\sigma$ is integer,
and we denote $\sigma=\pm S,\ S\in\Z_+$.
\begin{lemma}\label{remark3}
If $X^n\in\hat{R}$ for some $n$ such that $0\leq n<S$, then $\hat{R}=\hgc_1$.
In other words, any proper subalgebra $\hat{R}\subset\hgc_1$ is contained in
$\bigoplus_{n\geq S}\C X^n$.
\end{lemma}
\begin{proof}
If $n=0$, the claim follows immediately from Lemma \ref{remark1}.
Therefore we can assume $1\leq n< S$.
By (\ref{keq2m1}) we get $X^n _{<2n-1>} X^{n}\propto X^1$
and the coefficient is non zero since, by assumption, $\sigma^2>n^2$.
We then have that $X^1\in\hat{R}$.
By (\ref{keq3}) we get, for $m\geq2$, that
$X^1 _{<3>} X^m \propto X^{m-2}$
and the coefficient is non zero for all but at most one
value of $m\in\Z_+$ (the positive solution of the quadratic equation
$m^2-m+1=3S^2$).
By (\ref{keq3}) we have, for $m\geq3$ that
$X^1 _{<4>} X^m \propto X^{m-3}$ with non zero coefficient (since, by
assumption, $S>1$).
Putting together these two facts it is easy to understand that
the only possible situation is $\hat{R}=\hgc_1$.
\end{proof}
Assume $m$, $n\geq S$ where, as before, $\sigma=\pm S$.
It is immediate to check that the coefficient
$d^{(\sigma)}_{m,n,k}$ defined in (\ref{dmnk})
is positive if $m+n-k\geq S$ and it is zero otherwise.
In particular, under the condition $m+n-k\geq S$ all products
$X^m _{<k>} X^n$ with odd $k$ are non zero.
As immediate consequence we have the following:
\begin{lemma}\label{remark4}
{\bfseries (i)} For $\sigma=\pm S,\ S\in\Z_+$, there is a unique maximal
proper subalgebra of $\hgc_1$, namely
$$
\hrpms
=\text{span}_\C\{X^n \ ; \ n\geq S\} \ .
$$
{\bfseries (ii)} The following subspace
$$
\hrpmxs
=\text{span}_\C\{X^n \ ; \ n\geq S \ , \  n\in2\Z +1\}
$$
is the only other candidate to be a subalgebra of $\hgc_1$.
\end{lemma}
In fact we will see that $\hrpmxs$ is also
a subalgebra of $\hgc_1$.
But instead of showing it directly (which requires to prove
some non trivial identities of sums of binomial coefficients)
it will follow by Corollary \ref{baslem} after showing
that $\C[\partial]\hrpmxs$ is indeed a normalized subalgebra of $\gc_1$.


\begin{lemma}\label{subalg}
The following spaces:
\begin{eqnarray*}
\rps
&=& x^S\C[\partial,x] \ , \\
\rpxs
&=& \{x^S[p(\partial,x)+(-1)^{S+1} p(\partial,-\partial-x)]\ ,
\quad p(\partial,x)\in\C[\partial,x]\} \ , \\
\rms
&=& (x+\partial)^S\C[\partial,x] \ , \\
\rmxs
&=& \{(x+\partial)^S[p(\partial,x)+(-1)^{S+1} p(\partial,-\partial-x)]\ ,
\quad p(\partial,x)\in\C[\partial,x]\} \ ,
\end{eqnarray*}
are subalgebras of $\gc_1$.
They are normalized with respect to the Virasoro element
$\lpms=x+\frac{1\mp S}{2}\partial$.
\end{lemma}
\begin{proof}
The proof that $\rpms$ and $\rpmxs$ are closed under
$\lambda$--bracket is straightforward and can be found
in \cite{liberati}.
We need to prove that $\rpms$ and $\rpmxs$
are normalized subalgebras. In other words we need to show:
\begin{eqnarray}\label{**}
&& \lpms\ _{(i)}\ \rpms\ \subset\ \rpms \\
&& \lpms\ _{(i)}\ \rpmxs\ \subset\ \rpmxs \quad ,
\quad \text{ for } i=0,1,2 \ , \nonumber
\end{eqnarray}
where $\lpms=x+\frac{\mp S+1}{2}\partial$.
We already know that $\lpms\ _{(0)}\ A(\partial,y)=\partial A(\partial,y)$
and $\lpms\ _{(1)}\ A(\partial,y)=(n+1) A(\partial,y)$ if $A(\partial,y)$
is a homogeneous polynomial of degree $n$. Therefore (\ref{**})
is obviously satisfied for $i=0,1$, and we are left to prove that
$\rpms$ and $\rpmxs$
are invariant under the action of $\lpms\ _{(2)}$.
By the expression (\ref{xbracket}) of $\lambda$--bracket we have:
\begin{eqnarray}\label{***}
[\lps \ _\lambda \ x^S p(\partial,x)]
&=& (x+\partial+\frac{S+1}{2}\lambda)x^Sp(\partial+\lambda,x) \\
&-& (x+\frac{S-1}{2}\lambda)(-\lambda+x)^Sp(\partial+\lambda,-\lambda+x) \ .
\nonumber
\end{eqnarray}
To get the action of $\lpms\ _{(2)}$ we need to get the second derivative
with respect to $\lambda$ of both sides of (\ref{***}) and put
$\lambda=0$. After some algebraic manipulations one gets:
\begin{equation}\label{benoit}
\lps\ _{(2)}x^S p(\partial,x)
=x^S \{xD_2(2D_1-D_2)+\partial D_1^2+(s+1)(2D_1-D_2)\}p(\partial,x) \ ,
\end{equation}
where $D_1$ (resp. $D_2$) denotes the partial derivative with respect
to $\partial$ (resp. $x$).
From this expression we immediately get that $\rps$
is invariant under the action of $\lps\ _{(2)}$.
Also notice that the differential operator in parenthesis in the
right hand side of (\ref{benoit}) is invariant under the change of variables
$\partial\rightarrow\partial,\ x\rightarrow-\partial-x$.
This implies that $\rpxs$ is also invariant under the action
of $\lps\ _{(2)}$.
A similar calculation shows that $\rms$ and $\rmxs$ are invariant under the action of $\lms\ _{(2)}$.
\end{proof}
\begin{corollary}\label{novembre}
For every $S\in\Z_+$, we have:
\begin{eqnarray*}
\rpms &=& \C[\partial]\hrpms \ , \\
\rpmxs &=& \C[\partial]\hrpmxs \ .
\end{eqnarray*}
Moreover, given $L=x+\frac{1\mp S}{2}\partial$,
$\{\hrpms,\ \hrpmxs\}$ is a complete list
of infinite--dimensional subalgebras of $\hgc_1$
and $\{\rpms,\ \rpmxs\}$ is a complete list
of infinite rank normalized subalgebras of $\gc_1$.
\end{corollary}
\begin{proof}
By Corollary \ref{baslem}
there is a bijective correspondence between infinite--dimensional
subalgebras $\hat{R}\subset\hgc_1$ and infinite rank normalized subalgebras
$R\subset\gc_1$, given by $R=\C[\partial]\hat{R}$.
By Lemma \ref{remark4} there are at most two infinite--dimensional
subalgebras of $\hgc_1$, namely $\hrpms$ and $\hrpmxs$, and they satisfy
$\hrpmxs\subsetneq\hrpms$.
By Lemma \ref{subalg} there are at least two infinite rank normalized
subalgebras of $\gc_1$, namely $\rpms$ and $\rpmxs$,
and they satisfy: $\rpmxs\subsetneq\rpms$.
These two facts of course imply the claim.
\end{proof}


We can summarize all the results we have obtained:
\begin{theorem}\label{mathe}
{\bfseries (i)} Let $L=x+\frac{1-\sigma}{2}\partial$
be a Virasoro element of $\gc_1=\C[\partial,x]$.
Let $\hgc_1=\bigoplus_{n\in\Z_+}\C Q_n^{(\sigma)}(\partial,x)$ be the
corresponding reduced space.
For $\sigma\notin\Z$ the reduced space $\hgc_1$ has no proper
infinite--dimensional subalgebras.
In the case $\sigma\in\Z$, let $\sigma=\pm S$ with $S\in\Z_+$.
A complete list of infinite--dimensional proper
subalgebras of $\hgc_1$ is the following:
\begin{eqnarray*}
\hrpms
&=& \text{span}_\C\{Q_n^{(\sigma)}(\partial,x) \ ; \ n\geq S\} \\
\hrpmxs
&=&\text{span}_\C\{Q_n^{(\sigma)}(\partial,x) \ ; \ n\geq S,
n\in2\Z+1\} \ .
\end{eqnarray*}
{\bfseries (ii)} A complete list of infinite rank, normalized
(with respect to a Virasoro element), proper subalgebras
of $\gc_1$ is the following ($S\in\Z_+$):
\begin{eqnarray*}
\rps
&=& x^S\C[\partial,x] \ , \\
\rpxs
&=& \{x^S[p(\partial,x)+(-1)^{S+1} p(\partial,-\partial-x)] \ ,
\quad p(\partial,x)\in\C[\partial,x]\} \ , \\
\rms
&=& (x+\partial)^S\C[\partial,x]  \ , \\
\rmxs
&=& \{(x+\partial)^S[p(\partial,x)+(-1)^{S+1}
p(\partial,-\partial-x)] \ ,
\quad p(\partial,x)\in\C[\partial,x]\} \ ,
\end{eqnarray*}
where the corresponding Virasoro element is
$\lpms=x+\frac{1\mp S}{2}\partial$.
\end{theorem}
\begin{remark}
The subalgebras from Theorem \ref{mathe}(ii) that act irreducibly
on $\C[\partial]$ are $\rps$ and $\rpxs$.
\end{remark}
\begin{remark}
From Theorem \ref{mathe} we get, in particular, that
all infinite rank subalgebras of $\gc_1$ that contain a Virasoro element
are $R^{(\pm)}_0, R^{(\pm)}_1, R^{(\pm)}_{*,0}, R^{(\pm)}_{*,1}$.
\end{remark}


\ssection{Generalization to $\gc_N$}


\noindent
We want to generalize the results obtained in the previous section to
$\gc_N=\operatorname{Mat}_N\bC[\partial,x]$, $N\geq1$.
Our goal is to classify all conformal subalgebras $R\subset\gc_N$ which are
of infinite rank, which act irreducibly on $\C[\partial]^N$ and which are
normalized with respect to a given Virasoro element $L\in\gc_N$.
We will restrict ourselves to the case $L=(x+\alpha\partial)\id$,
$\alpha\in\C$.
As we pointed out earlier, this assumption is not restrictive only
for $N=1$.


First, we want to study the reduced space $\hgc_N$ and its algebra
structure.
Due to the fact that $L$ is proportional to the identity matrix,
all calculations done to find a basis for $\hgc_1$ can be repeated
for the reduced space $\hgc_N$. In other words
we have the following immediate generalization of
Theorem \ref{reducedspace}:
\begin{theorem}
The space $\hgc_N\subset\gc_N$, reduced with respect to the
Virasoro element $L=(x+\alpha\partial)\id$, is a vector space over $\bC$
with basis
$$
Q_n^{(\sigma)}(\partial,x) E_{ij} \qquad , \ n\geq0 \ , \ 1\leq i,j\leq N \ ,
$$
where $Q_n^{(\sigma)}(\partial,x)$ is defined by (\ref{qn}) and
$\sigma=1-2\alpha$.
\end{theorem}
By denoting, as before, $X^n=Q_n^{(\sigma)}(\partial,x)$, we can identify
the reduced space $\hgc_N$ with $\operatorname{Mat}_N\bC[X]$
(and the induced algebra structure).
To find the explicit expression of the products in $\hgc_N$, we
notice that the projection $\pi\ :\ \gc_N\rightarrow\hgc_N$
is simply obtained by:
$$
\pi(A(\partial,x))=A(0,X) \quad ,
\ A(\partial,x)\in\operatorname{Mat}_N\bC[\partial,x] \ .
$$
The proof of this relation is the same as the proof of
Lemma \ref{trivlemma}.
Therefore the same calculation leading to Theorem \ref{allprod}
gives the following:
\begin{theorem}\label{T}
The $k$-th product of elements of the reduced space
$\hgc_N=\operatorname{Mat}_N\bC[X]$ is given by:
\begin{equation}
X^mA _{<k>} X^nB
= [d^{(\sigma)}_{m,n,k}A\cdot B +(-1)^{k+1} d^{(-\sigma)}_{m,n,k}B\cdot A]
X^{m+n-k} \ ,
\end{equation}
where the coefficients $d^{(\sigma)}_{m,n,k}$ are given by (\ref{dmnk}).
\end{theorem}


In order to classify normalized subalgebras of $gc_N$ we use the
same considerations that we had for $\gc_1$. By Corollary \ref{baslem}
every normalized subalgebra $R\subset\gc_N$ corresponds,
via canonical projection, to a subalgebra of the reduced space:
$\hat{R}\subset\hgc_N$.
Furthermore $R$ is of infinite rank if and only if $\hat{R}$ is
infinite--dimensional over $\C$. For $N>1$ we also need to impose
the condition that $R$ acts irreducibly on $\C[\partial]^N$.
We will use the following:
\begin{lemma}\label{L}
(i) Any subalgebra $\hat{R}\subset\hgc_N=\operatorname{Mat}_N\bC[X]$
decomposes as:
\begin{equation}\label{1}
\hat{R}=\bigoplus_{n\geq0}X^nV_n \ ,
\end{equation}
where $V_n\subset\operatorname{Mat}_N\bC$ are linear subspaces. \\
(ii) If $R\subset\gc_N$ is a normalized subalgebra acting irreducibly
on $\C[\partial]^N$ and its projection $\hat{R}\subset\hgc_N$
decomposes as in (\ref{1}), then
$V=\sum_{n\geq0}V_n\subset\operatorname{Mat}_N\bC$
acts irreducibly on $\C^N$.
\end{lemma}
\begin{proof}
The proof of the first part is the same as the proof of
Lemma \ref{hrdec}.
For the second part, suppose $\hat{U}\subset\C^N$ is a subspace which
is invariant under the action of $V=\sum_{n\geq0}V_n$. We want
to prove that $U:=\C[\partial]\otimes\hat{U}$ is a proper submodule
of $R$, namely: $R_{(n)}U\subset U$, $\forall n\geq0$.
Given $a\in R$ and $v\in U$, we can decompose (uniquely)
$a=\sum_{i\geq0}\partial^ia^i$, with $a^i\in\hat{R}$ and
$v=\sum_{j\geq0}\partial^jv^j$, with $v^j\in\hat{U}$,
so that, by sesquilinearity, it suffices to prove:
$\hat{R}_{(n)}\hat{U}\subset U$, $\forall n\geq0$.
Let $a\in\hat{R}$ and $v\in\hat{U}$. By decomposition (\ref{1})
we can write:
$a=\sum_{n\geq0}Q_n^{(\sigma)}(\partial,x)A_n$,
where $A_n\in V_n$. The $\lambda$ action on $v$ is, by (\ref{xaction}):
\begin{eqnarray*}
a_\lambda v &=& \sum_{n\geq0}(Q_n^{(\sigma)}(\partial,x)A_n)_\lambda v \\
&=& \sum_{n\geq0}Q_n^{(\sigma)}(-\lambda,\lambda+\partial)A_n\cdot v
\in\C[\lambda,\partial]\otimes\hat{U}=\C[\lambda]\otimes U \ .
\end{eqnarray*}
This completes the proof.
\end{proof}


\begin{remark}\label{R}
As immediate consequence of Lemma \ref{L} and Theorem \ref{T}, a complete
classification of infinite--dimensional subalgebras $\hat{R}\subset\hgc_N$
such that $V$, defined in Lemma \ref{L}, acts irreducibly on $\C^N$
(or, equivalently, by Corollary \ref{baslem}, a complete classification
of infinite rank normalized subalgebras $R\subset\gc_N$ acting irreducibly
on $\C[\partial]^N$) is achieved once we find a list of all sequences
$\{V_n \ , \ n\in\Z_+\}$ of subspaces of $\operatorname{Mat}_N\C$
satisfying the following conditions:
\begin{enumerate}
\item $V=\sum_{n\geq0}V_n$ acts irreducibly on $\C^N$,
\item infinitely many of $V_n$'s are non zero,
\item if $A\in V_m$ and $B\in V_n$, then:
\begin{equation}\label{2}
[d^{(\sigma)}_{m,n,k}AB +(-1)^{k+1} d^{(-\sigma)}_{m,n,k}BA]\in V_{m+n-k} \ ,
\end{equation}
where $0\leq k\leq m+n$ and the coefficients $d^{(\sigma)}_{m,n,k}$ are
defined by (\ref{dmnk}).
\end{enumerate}
\end{remark}

\begin{remark}
The last condition on the spaces $V_n$'s takes a particularly simple form
for $k=0,1$. By (\ref{dmnk}) we have:
$$
d^{(\sigma)}_{m,n,0} = 1 \ , \quad \forall m,n\geq0 \ \ \text{ and } \
d^{(\sigma)}_{m,n,1} = \frac{1}{2}(m+n) \ , \quad \forall m,n\geq1 \ ,
$$
so that (\ref{2}) implies the following condition:

\vspace{.1in}

$3^\prime$. If $A\in V_m$, $B\in V_n$, then:
\begin{eqnarray}\label{y1}
\left[A,B\right]_- &\in& V_{m+n} \ , \quad \forall m,n\geq0 \nonumber\\
\left[A,B\right]_+ &\in& V_{m+n-1} \ , \quad \forall m,n\geq1 \ .
\end{eqnarray}
Here and further, $[A,B]_\pm:=AB\pm BA$, as usual.
\end{remark}

\ssubsection{Preliminary results}

\noindent
We want to classify all sequences $\{V_n\ ,\ n\in\Z_+\}$ satisfying
conditions 1--3 of Remark \ref{R}.
We will use the following notation (assuming $V_{-1}=0$):
\begin{eqnarray}\label{y2}
V^{(i)}_+ &=& \sum_{n\geq i}V_{2n-1} \quad , \quad
V^{(i)}_- \ \ =\ \ \sum_{n\geq i}V_{2n} \ ,\nonumber\\
V^{(i)} &=& V^{(i)}_++V^{(i)}_- = \sum_{n\geq 2i-1}V_{n} \ ,
\end{eqnarray}
where $i\in\Z_+$. In this section we classify all possibilities for
the spaces $V^{(i)}_\pm$. The final result is:
\begin{proposition}\label{P1}
The spaces $V^{(i)}_\pm$ are all equal, independently on $i$:
$$
V^{(i)}_+ = V_+ \quad,\quad V^{(i)}_- = V_- \quad \forall i\in\Z_+ \ ,
$$
and the only allowed possibilities for $V_\pm$ are:
\begin{enumerate}
\item $V_+=V_-=\operatorname{Mat}_N\C$,
\item $V_\pm=\{A\in\operatorname{Mat}_N\C\ |\ A^*=\pm A\}$,
where $*$ is any linear antiinvolution on $\operatorname{Mat}_N\C$,
namely a linear map
$*:\ \operatorname{Mat}_N\C\rightarrow\operatorname{Mat}_N\C$
such that: $A^{**}=A$ and $(AB)^*=B^*A^*$ for all $A,B\in\operatorname{Mat}_N\C$.
\end{enumerate}
In particular
$$
V^{(i)}=\operatorname{Mat}_N\C \ , \quad \forall i\in\Z_+ \ .
$$
\end{proposition}
The proof of this proposition is based on several lemmas.
\begin{lemma}\label{L1}
For each $i\geq1$ the space $V^{(i)}$ is isomorphic to a matrix algebra:
$V^{(i)}\simeq\operatorname{Mat}_{M_i}\C$
for some positive integer $M_i\leq N$.
\end{lemma}
\begin{proof}
By relations (\ref{y1}) it follows that
$[V,V]_-\subset V$, so that $V$ is
a Lie subalgebra of $\operatorname{gl}_N$ acting irreducibly on $\C^N$.
By Cartan--Jacobson Theorem, $V$ is then either semisimple or semisimple
plus scalars.
By (\ref{y1}) we also have
$[V,V^{(i)}]_-\subset V^{(i)}$,
so that $V^{(i)}$ is an ideal of $V$; the only possibilities are
either: $V^{(i)}=\bigoplus($ simple components of $V)$,
or: $V^{(i)}=\bigoplus($ simple components of $V) \oplus\C\id$.
In particular, for each $i\geq1$, $V$ decomposes as a
direct sum of Lie algebras:
\begin{equation}\label{y3}
V=V^{(i)}\oplus\operatorname{g}^{(i)} \ ,
\end{equation}
where $\operatorname{g}^{(i)}$ is semisimple and
the center
of $V^{(i)}$ is either zero or $\C\id$.
Again by (\ref{y1}) we have
$[V^{(i)},V^{(i)}]_\pm\subset V^{(i)}$,
so that $V^{(i)}$ is an associative subalgebra of $\operatorname{Mat}_N\C$.
Let us denote by $I\subset V^{(i)}$ the nilradical of $V^{(i)}$ (viewed
as associative algebra). By definition $I$ is a two--sided ideal of
the associative algebra $V^{(i)}$; in particular it is an ideal
of $V^{(i)}$, viewed as Lie algebra, and it is invariant under the action
of any automorphism of the Lie algebra $V$.
This guarantees that $I$ is an ideal of the Lie algebra $V$.
But then $I$ is an ideal of $V$ consisting of nilpotents elements,
and, since $V$ is semisimple or semisimple plus scalars,
we conclude that $I=0$. In other words $V^{(i)}$ is a semisimple
associative algebra, so that it decomposes as a direct sum of
matrix algebras:
$$
V^{(i)}\simeq\bigoplus_k\operatorname{Mat}_{M_k}\C \ .
$$
But then if we view $V^{(i)}$ as a Lie algebra, its center will have
dimension equal to the number of matrix algebras
in this decomposition, so that, by the previous result,
$V^{(i)}$ must be isomorphic to a single matrix algebra:
$V^{(i)}\simeq\operatorname{Mat}_{M_i}\C$ for some $M_i\leq N$.
\end{proof}
We can now prove the following stronger result:
\begin{lemma}\label{L2}
$$
V^{(i)}=V=\operatorname{Mat}_N\C \ , \quad \forall i\in\Z_+ \ .
$$
\end{lemma}
\begin{proof}
By definition $V^{(i+1)}\subset V^{(i)}$ for all $i\in\Z_+$ and, by
the second condition in Remark \ref{R}, all spaces $V^{(i)}$ are
non zero.
We then have a non increasing sequence of positive integers:
$N\geq M_1\geq M_2 \geq\cdots>0$ such that:
$V^{(i)}\simeq\operatorname{Mat}_{M_i}\C$.
Such a sequence eventually will stabilize, namely for some $I>>0$
we have:
$$
N\geq M_i=\bar{M}>0 \quad , \quad
\operatorname{Mat}_N\C\supset V^{(i)}
=\bar{V}\simeq\operatorname{Mat}_{\bar{M}}\C
\quad , \quad \forall i\geq I \ .
$$
It is clear that, in order to prove the Lemma, it suffices
to prove: $\bar{M}=N$.
Recall that by (\ref{y3}) the Lie algebra
$V=\sum_{n\geq0}V_n\subset\operatorname{gl}_N$
decomposes as a direct sum of Lie algebras:
$V=\bar{V}\oplus\bar{\operatorname{g}}$
where $\bar{V}$, by Lemma \ref{L1}, is an ideal isomorphic to
$\operatorname{gl}_{\bar{M}}$ for some $\bar{M}\leq N$, while
$\bar{\operatorname{g}}$ is a complementary ideal and it
is a semisimple Lie algebra.
%
%
%
%
%
%
%
%
%
%
%
%
As a Lie algebra, $\bar{V}\simeq\operatorname{gl}_{\bar{M}}$ has a one
dimensional center which, by (\ref{y3}),
has to be equal to $\C\id_N$.
Hence $\id_N\in\bar{V}$.
By (\ref{y1}) and (\ref{y2}) it follows, for $i\geq I$:
$$
\operatorname{g} = [\id_N,\operatorname{g}]_+
\subset[\bar{V},V]_+ = [V^{(i+1)},V]_+ \subset V^{(i)} =\bar{V}
$$
and this is possible only if $\operatorname{g}=0$.
In conclusion we have
$V=\bar{V}\simeq\operatorname{Mat}_{\bar{M}}\C$,
which clearly implies $\bar{M}=N$ (since, by assumption, $V$ acts irreducibly
on $\C^N$).
\end{proof}
We can now consider the spaces $V^{(i)}_\pm$.
By definition (\ref{y2}) and conditions (\ref{y1}) it is immediate to check
that for $i\geq1$:
$$
[V^{(i)}_a,V^{(i)}_b]_c\subset V^{(i)}_{abc} \quad , \quad a,b,c=\pm \ ,
$$
and, by Lemma \ref{L2} we also have:
$$
V^{(i)}_++V^{(i)}_-=\operatorname{Mat}_N\C \ .
$$
There are only two possibilities for spaces $V^{(i)}_\pm$ satisfying
the above conditions. This is stated in the following:
\begin{lemma}\label{L3}
Let $\{V_+,V_-\}$ be a pair of linear subspaces of $\operatorname{Mat}_N\C$
such that:
\begin{eqnarray}\label{X}
V_+\cap V_- &\neq& \operatorname{Mat}_N\C \ , \nonumber\\
V_++V_- &=& \operatorname{Mat}_N\C \ , \\
\left[V_a,V_b\right]_c &\subset& V_{abc}\ ,\ \forall a,b,c=\pm\ . \nonumber
\end{eqnarray}
Then there is a (unique) linear antiinvolution
$*\ : \operatorname{Mat}_N\C\rightarrow\operatorname{Mat}_N\C$
such that:
\begin{equation}\label{5}
V_\pm=\{A\in\operatorname{Mat}_N\C\ |\ A^*=\pm A\} \ .
\end{equation}
Conversely, given any antiinvolution $*$ of $\operatorname{Mat}_N\C$,
the subspaces $V_\pm$ defined by (\ref{5}) satisfy all conditions (\ref{X}).
\end{lemma}
\begin{proof}
Let $*$ be a linear antiinvolution of $\operatorname{Mat}_N\C$ and
define $V_\pm$ as in (\ref{5}). It is clear that
$\operatorname{Mat}_N\C=V_+\oplus V_-$.
Furthermore, let $a,b,c=\pm$ and suppose $A\in V_a$, $B\in V_b$. We have:
$$
(\left[A,B\right]_c)^*=(AB+cBA)^*=abc(AB+cBA)=abc\left[A,B\right]_c^* \ ,
$$
which proves that $V_\pm$ satisfy conditions (\ref{X}).
Conversely, let $V_\pm$ be subspaces of $\operatorname{Mat}_N\C$
satisfying (\ref{X}).
We want to construct the corresponding antiinvolution
$*:\ \operatorname{Mat}_N\C\rightarrow\operatorname{Mat}_N\C$.
For this, notice that
$$
[V_++V_-,V_+\cap V_-]_\pm\subset V_+\cap V_-\ ,
$$
so that $V_+\cap V_-$ is an ideal of $\operatorname{Mat}_N\C\simeq V_++V_-$
with respect to both the Lie algebra structure and the Jordan algebra
structure.
Since $\operatorname{Mat}_N\C$ is a simple associative algebra,
it follows that
$V_+\cap V_-=0$ or $=\operatorname{Mat}_N\C$.
Since by (\ref{X}) $V_+\cap V_-\neq\operatorname{Mat}_N\C$,
we have $V_+\cap V_-=0$, which implies:
$\operatorname{Mat}_N\C=V_+\oplus V_-$.
In this case the corresponding linear map
$*:\ \operatorname{Mat}_N\C\rightarrow\operatorname{Mat}_N\C$
is uniquely defined by the conditions:
$A^*=\pm A$ for $A\in V_\pm$, and it is easy to check that
it is an antiinvolution of $\operatorname{Mat}_N\C$.
\end{proof}
We can now complete the proof of Proposition \ref{P1}. In order to do so
we are left to prove:
\begin{lemma}
All spaces $V^{(i)}_\pm$ are equal (i.e. independent on $i$).
\end{lemma}
\begin{proof}
By definition we have:
$V^{(0)}_\pm\supset V^{(1)}_\pm \supset V^{(2)}_\pm\supset \cdots$,
so that at some point the sequence will stabilize, namely there is
$I>>0$ such that:
$V^{(i)}_\pm=\bar{V}_\pm$, for all $i\geq I$.
Since for each $j\geq0$ we have:
$V^{(0)}_\pm\supset V^{(j)}_\pm \supset \bar{V}_\pm$, we just need to
show: $V^{(0)}_\pm = \bar{V}_\pm$.
For this we use Lemma \ref{L3}.
Clearly if
$V^{(0)}_\pm=\{A\in\operatorname{Mat}_N\bC\ |\ A^*=\pm A\}
\nsubseteq\operatorname{Mat}_N\bC$,
then $V^{(0)}_\pm\subset\bar{V}_\pm$ which implies
$V^{(0)}_\pm=\bar{V}_\pm$.
Therefore we are left to consider the case:
$V^{(0)}_+=V^{(0)}_-=\operatorname{Mat}_N\bC$. We want to prove in this
case that $\bar{V}_+=\bar{V}_-=\operatorname{Mat}_N\bC$.
Since $\id_N\in\bar{V}_+$, we have, for $i\geq I$:
\begin{eqnarray*}
\operatorname{Mat}_N\bC
&=& \left[\operatorname{Mat}_N\bC,\id_N\right]_+
\subset \left[V^{(0)}_-,V^{(i+1)}_+\right]_+ \\
&\subset& \sum_{m\geq0, \atop{n\geq i+1}}\left[V_{2m},V_{2n-1}\right]_+
\subset \sum_{m\geq0, \atop{n\geq i+1}}V_{2(m+n)-2}
\subset V^{(i)}_-=\bar{V}_-
\end{eqnarray*}
so that: $\bar{V}_-=\operatorname{Mat}_N\bC$ and then, by Lemma \ref{L3},
$\bar{V}_+=\operatorname{Mat}_N\bC$ as well, thus
proving the claim.
\end{proof}

\ssubsection{Case $\sigma\not\in\Z$}

\noindent
We will first consider the case $\sigma\not\in\Z$.
The basic observation is that the coefficients $d^{(\sigma)}_{m,n,k}$
defined in (\ref{dmnk}) take, for $m=n,\ k=2n-1$, the simple form:
$d^{(\sigma)}_{n,n,2n-1} = \frac{n+1}{\binom{2n}{n}}
\prod_{i=2}^{n}(i^2-\sigma^2)$,
so that, for $\sigma$ not integer, we have
$d^{(\sigma)}_{n,n,2n-1}=d^{(-\sigma)}_{n,n,2n-1} \neq0$
and (\ref{2}) implies the following:
\begin{lemma}\label{D1}
If $A,B\in V_n$ for some $n\geq1$ then: $[A,B]_+\in V_1$.
\end{lemma}
As a consequence of Lemma \ref{D1} we can prove the following:
\begin{lemma}\label{L4}
$\id\in V_1$.
\end{lemma}
\begin{proof}
First notice that, if for some $n\geq1$ there is a nondegenerate matrix
$A\in V_n$, then $B:=A^2$ is non degenerate and, by Lemma \ref{D1},
$B, B^2, B^3, \cdots\in V_1$, so that, by the Cayley--Hamilton Theorem
$$
\id\in\text{span}_\C\{B,B^2,\cdots,B^N\}\subset V_1 \ ,
$$
and the claim is proved.
Given a matrix $A\in\operatorname{Mat}_N\C$, we denote by $U_A[0]$
 the generalized eigenspace of $A$ with eigenvalue zero. Let:
$$
k:=\text{ min } \{\text{ dim }U_A[0]\ |\ A\in V_n\ ,\ n\geq1\} \ .
$$
For $k=0$ the Lemma has already been proved. On the other hand $k<N$ since,
by Proposition \ref{P1},
$\id\in\operatorname{Mat}_N\C=\sum_{n\geq1}V_n$
and it is not possible to write the identity matrix as a linear combination
of nilpotent matrices.
We are left to consider the case $0<k<N$. Let $A\in V_n$, $n\geq1$,
be a matrix such that dim $U_A[0]=k$. By Lemma \ref{D1} we have
$A^2\in V_1$ and clearly dim $U_{A^2}[0]= $dim$ U_A[0]=k$, so that
we can assume, without loss of generality, that $A\in V_1$.
After a change of basis $A$ can be written in Jordan form:
$$
A=T\left[\begin{array}{ll}
\Lambda^\prime & 0 \\
0 & \Lambda_0
\end{array}\right]T^{-1}
$$
where $\Lambda^\prime$ is a non degenerate upper triangular
$(N-k)\times(N-k)$ matrix and $\Lambda_0$ is a nilpotent strictly
upper triangular $k\times k$ matrix.
By Lemma \ref{D1} we know that $A,A^2,A^3,\cdots\in V_1$ and, since
$\Lambda^\prime$ is non degenerate and $\Lambda_0$ is nilpotent, we get
using the Cayley--Hamilton Theorem that:
$$
P=T\left[\begin{array}{ll}
\id & 0 \\
0 & 0
\end{array}\right]T^{-1} \in V_1 \ .
$$
(After taking a sufficiently large power, $\Lambda_0$ becomes zero, and
after taking a suitable polynomial of the resulting matrix $A^n$,
$\Lambda^\prime$ becomes the identity matrix.)
Now let $C\in V_n$ for some $n\geq1$. We can write it as:
$$
C=T\left[\begin{array}{ll}
C^\prime & C_1 \\
C_2 & C_0
\end{array}\right]T^{-1} \ ,
$$
where $C^\prime$ is $(N-k)\times(N-k)$ and $C_0$ is $k\times k$.
Since $\sum_{n\geq1}V_n=\operatorname{Mat}_N\C$ we can choose $C$
so that $C_0$ is not nilpotent (i.e.
dim $U_{C_0}[0]<k$).
The second equation of (\ref{y1}) guarantees:
$$
\tilde{C}=C-[P,C]_+
=T\left[\begin{array}{ll}
-C^\prime & 0 \\
0 & C_0
\end{array}\right]T^{-1}
 \in V_n\ .
$$
But then, by Lemma \ref{D1} we get that:
$$
D=\tilde{C}^2
=T\left[\begin{array}{ll}
D^\prime & 0 \\
0 & D_0
\end{array}\right]T^{-1}
\in V_1
$$
and since $C_0$ is not nilpotent, so is $D_0=C_0^2$.
We then proved that both $P$ and $D$ are in $V_1$, and, since
dim $U_{D_0}[0]<k$, we can always find a linear combination
$E=D+\alpha P\in V_1$ such that dim $U_E[0]<k$.
This contradicts the very definition of $k$ and shows that
the case $k>0$ is not allowed.
\end{proof}
Using the previous result we can finally prove:
\begin{lemma}
$V_n=\operatorname{Mat}_N\C\ ,\ \forall n\geq0$.
\end{lemma}
\begin{proof}
By looking at the coefficients $d^{(\sigma)}_{m,n,k}$ defined in (\ref{dmnk})
for the particular case of $m=1$, $k=3$, we have:
$$
d^{(\sigma)}_{1,n,3}+d^{(-\sigma)}_{1,n,3}
=\frac{n(n-1)}{2(2n-1)}((n^2-n+1)-3\sigma^2) \ ,
$$
which is non zero
as soon as $n\geq I$, where $I\in\Z_+$ is sufficiently large.
Since $\id\in V_1$, condition (\ref{2}) implies, by taking $m=1$ and $k=3$,
that: $V_n \subset V_{n-2}$.
We then have by Proposition \ref{P1}
and a simple induction argument:
$$
V_{2n-1}=V_+ \quad , \quad V_{2n}=V_- \quad , \quad \forall n\geq I \ .
$$
Now let us apply (\ref{2}) with $m=1$, $n\geq I$ and $k=4$. After
a straightforward calculation, one gets by (\ref{dmnk}):
\marginpar{checked with maple}
$$
d^{(\sigma)}_{1,n,4}+d^{(-\sigma)}_{1,n,4}
= -\sigma(\sigma^2-1)(n-2)\neq0 \ .
$$
Using again Lemma \ref{L4} we thus get $V_-=V_{2n}\subset V_{2n-3}=V_+$,
and this is possible only if
$V_+=V_-=\operatorname{Mat}_N\C$. We therefore proved that
$V_n=\operatorname{Mat}_N\C$ for sufficiently
large $n$.
Hence for $n$ large enough: $\id\in V_n, V_{n+1}$,
so that:
$$
(d^{(\sigma)}_{n,n+1,2n+1}+d^{(-\sigma)}_{n,n+1,2n+1})\id\in V_0 \ ,
$$
and since, by a computation:
$d^{(\sigma)}_{n,n+1,2n+1}
=\frac{n+1-\sigma}{2\binom{2n}{n}}\prod_{k=1}^n(k^2-\sigma^2)$,
we have that $d^{(\sigma)}_{n,n+1,2n+1}+d^{(-\sigma)}_{n,n+1,2n+1}\neq0$
for $\sigma\not\in\Z$, so that $\id\in V_0$.
To conclude we just notice that, since $\id\in V_0$, condition (\ref{2})
guarantees:
$$
A\in V_n \ \Rightarrow \
(d^{(\sigma)}_{0,n,1}+d^{(-\sigma)}_{0,n,1})A\in V_{n-1} \ .
$$
Since, by a computation:
$d^{(\sigma)}_{0,n,1}+d^{(-\sigma)}_{0,n,1}=n\neq0$, we have
$V_n\subset V_{n-1}$ for $n\geq1$, which clearly implies the claim.
\end{proof}
We summarize the above results in the following:
\begin{proposition}
For $\sigma\not\in\Z$ the only sequence $\{V_n \ , \ n\in\Z_+\}$ of subspaces
of $\operatorname{Mat}_N\C$ satisfying conditions 1--3 of
Remark \ref{R} is given by: $V_n=\operatorname{Mat}_N\C \ , \
\forall n\in\Z_+$.
\end{proposition}

\ssubsection{Case $\sigma\in\Z$}

\noindent
We are left to consider the situation $\sigma\in\Z$. We will denote,
as in Section 3, $\sigma=\pm S,\ S\in\Z_+$ and we will let
$\bar{S}=S$ (resp. $=S+1$) if $S$ is odd (resp. even).
As we already pointed out (immediately after Lemma \ref{remark3}),
for $m, n\geq S$ the coefficients
$d^{(\sigma)}_{m,n,k}$ defined by (\ref{dmnk}) are positive for $m+n-k\geq S$
and zero otherwise.
We then get that (\ref{2}) with $k=2n-\bar{S}\in2\Z_++1$
implies the following:
\begin{lemma}
Let $n\geq S$. If $A\in V_n$, then $A^2\in V_{\bar{S}}$.
\end{lemma}
We can then use arguments similar to the ones used to prove Lemma \ref{L4}
to get:
\begin{lemma}
$\id\in V_{\bar{S}}$.
\end{lemma}
\begin{proof}
Let $k:=$ min $\{$ dim $U_A[0]\ |\ A\in V_n\ ,\ n\geq\bar{S}\}$, where again
$U_A[0]$ denotes the generalized eigenspace of $A$ with eigenvalue zero.
$k=N$ is not allowed since it contradicts
$\sum_{n\geq\bar{S}}V_n=\operatorname{Mat}_N\C$, and for $k=0$ the claim
follows by the Cayley--Hamilton Theorem.
Suppose then $0<k<N$.
Using (\ref{2}) and the Cayley--Hamilton Theorem we get, as in the proof
of Lemma \ref{L4}, that:
$$
P=T\left[\begin{array}{ll}
\id & 0 \\
0 & 0
\end{array}\right]T^{-1} \in V_{\bar{S}}
$$
for some invertible matrix $T$. Here dim Ker$(P)=k$.
We will consider separately the two cases $\bar{S}=1$
and $\bar{S}>1$.

For $\bar{S}=1$ (namely $\sigma=0$ or $\pm1$) we will proceed
as in the proof of Lemma \ref{L4}.
Let $C\in V_n$, $n\geq\bar{S}$. We can write it as:
$$
C=T\left[\begin{array}{ll}
C^\prime & C_1 \\
C_2 & C_0
\end{array}\right]T^{-1} \in V_n \ ,
$$
where $C_0$ is a $k\times k$ matrix and we may assume,
without loss of generality, that it is not nilpotent.
By (\ref{2}) we then have:
$d^{(\sigma)}_{\bar{S},n,\bar{S}} PC +d^{(-\sigma)}_{\bar{S},n,\bar{S}} CP
\in V_n$. For $\bar{S}=1$ we already know that
\begin{equation}\label{M4}
d^{(\sigma)}_{\bar{S},n,\bar{S}}=d^{(-\sigma)}_{\bar{S},n,\bar{S}}\neq0 \ ,
\end{equation}
so that we deduce:
$$
\tilde{C}=C-[P,C]_+=
T\left[\begin{array}{ll}
-C^\prime & 0 \\
0 & C_0
\end{array}\right]T^{-1} \in V_n \ ,
$$
and, as for Lemma \ref{L4}, we can conclude from this that $k>0$
is not allowed.
Notice that for $\bar{S}>1$ this argument works provided that
(\ref{M4})holds for every value of $\bar{S}$.
This is actually true, but the proof of it is not trivial and involves
rather complicated identities of sum and products of binomial coefficients.
Instead we will use, for $\bar{S}>1$, a different argument.
Then the validity of (\ref{M4}) for $\bar{S}>1$ will follow as a corollary
of Theorem \ref{finthe}.

In the case $\bar{S}>1$ we first notice that $P\in V_n$ for every
odd integer $n$ such that $n\geq\bar{S}$.
This follows by (\ref{2}) and the fact that $P$ is a projection.
We then have that, for some odd $n\geq\bar{S}$,
$$
P=T\left[\begin{array}{ll}
\id & 0 \\
0 & 0
\end{array}\right]T^{-1}
\quad , \quad
C=T\left[\begin{array}{ll}
C^\prime & C_1 \\
C_2 & C_0
\end{array}\right]T^{-1} \in V_n \ ,
$$
with $C_0$ not nilpotent, so that, by (\ref{2}):
$$
d^{(\sigma)}_{n,n,n}PC+d^{(-\sigma)}_{n,n,n}CP \ \in V_n \ ,
$$
and, since obviously:
$d^{(\sigma)}_{n,n,n}=d^{(-\sigma)}_{n,n,n}>0$,
this means $[P,C]_+\in V_n$.
We then have: $\tilde{C}=C-[P,C]_+\in V_n$. From here we can
repeat the same argument as before to conclude the proof.
\end{proof}
Using the previous result we can show:
\begin{lemma}
$V_n=V_{(-1)^{n+1}}\qquad \forall n\geq S$.
\end{lemma}
\begin{proof}
By (\ref{2}) we have that, for $A\in V_n, \ n\geq S+2$:
$$
(d^{(\sigma)}_{\bar{S},n,\bar{S}+2}+d^{(-\sigma)}_{\bar{S},n,\bar{S}+2})
A\in V_{n-2}
$$
Since both coefficients $d^{(\sigma)}_{\bar{S},n,\bar{S}+2}$ and
$d^{(-\sigma)}_{\bar{S},n,\bar{S}+2}$ are positive, we deduce that
$V_n\subset V_{n-2}$ for any $n\geq S+2$. This, combined to the fact that
$V_+=\sum_{n\geq S}V_{2n-1}$, $V_-=\sum_{n\geq S}V_{2n}$, gives
by induction the claim.
\end{proof}
We are left to consider the spaces $V_n$ with $n<S$.
We can assume in the following that $S\geq1$.
The solution is stated in the following:
\begin{lemma}
{\bf (i)} All spaces $V_n$ with $n<S$ are equal. We will denote:
$J=V_n$ for $0\leq n<S$. For $\sigma=S$ (resp. $\sigma=-S$), $J$ is a
left (resp. right) ideal of the associative algebra
$\operatorname{Mat}_N\bC$, i.e.:
$\operatorname{Mat}_N\bC\cdot J\subset J$
(resp. $J\cdot\operatorname{Mat}_N\bC\subset J$). \\
{\bf (ii)} Furthermore, if $J\neq0$, then necessarily:
$V_+=V_-=\operatorname{Mat}_N\bC$.
\end{lemma}
\begin{proof}
For the proof we will use the following simple facts about the coefficients
$d^{(\sigma)}_{m,n,k}$:
\begin{enumerate}
\item $d^{(S)}_{m,n,k}=0$, if $0\leq m<S,\ n\geq S,\ k>m+n-S$;
\item in general $d^{(-S)}_{m,n,k}$ can be zero, but it is certainly
non zero for some specific values of $k$:
\begin{enumerate}
\item $d^{(-S)}_{m,n,m+n}\neq0$, if $0\leq m<S,\ n\geq S$;
\item $d^{(-S)}_{0,n,k}\neq0$, if $n-S<k\leq n$.
\end{enumerate}
\end{enumerate}
It follows from (\ref{2}) and the above facts that:
\begin{enumerate}
\item if $A\in V_m,\ B\in V_n$ for $0\leq m<S,\ n\geq S$, then:
\begin{eqnarray}\label{M1}
\text{for } \sigma=+S\ &:& \ BA\in V_0 \\
\text{for } \sigma=-S\ &:& \ AB\in V_0 \nonumber
\end{eqnarray}
\item if $A\in V_0,\ B\in V_n$ with $n\geq S$, then:
\begin{eqnarray}\label{M2}
\text{for } \sigma=+S\ &:& \ BA\in V_m \quad
\forall m \text{ such that } 0\leq m<S \\
\text{for } \sigma=-S\ &:& \ AB\in V_m \quad
\forall m \text{ such that } 0\leq m<S \nonumber
\end{eqnarray}
\end{enumerate}
From (\ref{M1}) we immediately deduce that $V_0=\sum_{0\leq n<S}V_n$
and it is a left (resp. right) ideal of $\operatorname{Mat}_N\C$
for $\sigma=+S$ (resp. $\sigma=-S$).
From (\ref{M2}) we get that $V_m=V_0,\ \forall m<S$, thus proving the first
part of the lemma.

For the second part assume $V_n=J\neq0$ for every $n$ such that $0\leq n<S$.
Let $A\in V_0$ be a non zero matrix. Since $\id\in V_n$ for every odd
integer $n\geq S$, we get from (\ref{2}), by choosing $k=1,2$, that:
\begin{eqnarray}\label{M3}
&& (d^{(\sigma)}_{0,n,1}+d^{(-\sigma)}_{0,n,1})A\in V_{n-1} \nonumber\\
&& (d^{(\sigma)}_{0,n,2}-d^{(-\sigma)}_{0,n,2})A\in V_{n-2} \ .
\end{eqnarray}
By explicit computation
we have:
$d^{(\sigma)}_{0,n,1}+d^{(-\sigma)}_{0,n,1}=n\neq0$
and
$d^{(\sigma)}_{0,n,2}-d^{(-\sigma)}_{0,n,2}=-(n-1)\sigma \neq0$,
so that we have $0\neq A\in V_{n-1}\cap V_{n-2}=V_+\cap V_-$
and this is possible, by Proposition \ref{P1}, only if
$V_+=V_-=\operatorname{Mat}_N\C$.
\end{proof}
We can summarize the results obtained so far in the following:
\begin{proposition}\label{PF}
Suppose $\sigma\in\Z$ and denote $\sigma=\pm S,\ S\in\Z_+$.
If a sequence $\{V_n\ ,\ n\geq0\}$
of subspaces of $\operatorname{Mat}_N\C$ satisfies conditions 1--3
of Remark \ref{R}, then it must be either:
\begin{eqnarray*}
V_n &=& J \qquad \text{for } 0\leq n<S \ , \\
V_n &=& \operatorname{Mat}_N\C \qquad \text{ for } n\geq S \ ,
\end{eqnarray*}
where $J$ is a left (resp. right) ideal of $\operatorname{Mat}_N\C$
if $\sigma=+S$ (resp. $\sigma=-S$), or:
\begin{eqnarray*}
V_n &=& 0 \qquad \text{ for } 0\leq n<S \ , \\
V_n &=& V_{(-1)^{n+1}} \quad \text{ for } n\geq S \ ,
\end{eqnarray*}
where $V_\pm=\{A\in\operatorname{Mat}_N\C\ |\ A^*=\pm A\}$
for some linear antiinvolution $*$ of $\operatorname{Mat}_N\C$.
\end{proposition}
\begin{remark}
We did not prove so far that all the above sequences $\{V_n\ ,\ n\geq0\}$
satisfy conditions 1--3 of Remark \ref{R}.
A direct proof of it would involve non trivial identities between sums of
multinomial coefficients. Instead we will prove this fact by looking
at the corresponding normalized subalgebras of $\gc_N$, and we will deduce
from that some interesting combinatorial identities.
\end{remark}

\ssubsection{Classification of normalized subalgebras of $\hgc_N$}

\noindent
According to Lemma \ref{L} and Remark \ref{R} we can translate the
results obtained in the previous Sections in term of classification of
infinite--dimensional subalgebras of the reduced space $\hgc_N$:
\begin{corollary}\label{kari}
{\bf (i)} For $\sigma\not\in\Z$ there are no proper infinite--dimensional
subalgebras of $\hgc_N$. \\
{\bf (ii)} For $\sigma=\pm S$, $S\in\Z_+$, the only
candidates to be infinite--dimensional, proper subalgebras of $\hgc_N$
acting irreducibly on $\C^N$ are:
$$
\hrpmsj
= \left(\bigoplus_{0\leq n<S}X^n J\right)\oplus
\left(\bigoplus_{n\geq S}X^n\operatorname{Mat}_N\bC\right) \ ,
$$
where $J\subset\operatorname{Mat}_N\C$ is a left (resp. right) ideal
of $\operatorname{Mat}_N\C$ for $\sigma=+S$ (resp. $\sigma=-S$), and:
\begin{equation}\label{rx}
\hrpmxs
=\bigoplus_{n\geq S}X^n V_{(-1)^{n+1}}
\end{equation}
where $V_\pm=\{A\in\operatorname{Mat}_N\C\ |\ A^*=\pm A\}$
and $*$ is a linear antiinvolution of $\operatorname{Mat}_N\C$.
\end{corollary}
\begin{remark}
Any left (resp. right) ideal of $\operatorname{Mat}_N\C$ is principal,
namely it is of the form $J=(\operatorname{Mat}_N\C) A$
(resp. $J=A(\operatorname{Mat}_N\C)$), for some matrix
$A\in\operatorname{Mat}_N\C$.
A special case is when we take the matrix $A$ to be:
\begin{equation}\label{ikn}
I_{k,N}=
\left[\begin{array}{cc}
\id & 0 \\
0 & 0
\end{array}\right]
\end{equation}
of rank $k$. The left (resp. right) ideal generated by $I_{k,N}$
is $J^+_k=(\operatorname{Mat}_N\C) I_{k,N}$
(resp. $J^-_k=I_{k,N}(\operatorname{Mat}_N\C)$).
In the following we will denote:
$$
\hrpmsk=\hrpmsjpmk \quad,\qquad k=0,\dots,N\ ,
$$
namely:
\begin{eqnarray}\label{rk}
\hrpsk
&=& \left(\bigoplus_{0\leq n<S}X^n \operatorname{Mat}_N\bC\ I_{k,N}\right)\oplus
\left(\bigoplus_{n\geq S}X^n\operatorname{Mat}_N\bC\right) \ , \nonumber\\
\hrmsk
&=& \left(\bigoplus_{0\leq n<S}X^n I_{k,N}\operatorname{Mat}_N\bC\right)\oplus
\left(\bigoplus_{n\geq S}X^n\operatorname{Mat}_N\bC\right) \ .
\end{eqnarray}
\end{remark}
We now want to find the corresponding normalized
subalgebras of $\gc_N$.
We will use an argument similar to the one used for $\gc_1$.
\begin{lemma}\label{wailin}
The following spaces ($S\in\Z_+$):
\begin{eqnarray}\label{wa5}
\rpsk
&=& \operatorname{Mat}_N\C[\partial,x](I_{k,N}+x^S\bar{I}_{k,N})\ ,\nonumber\\
\rpxs
&=& \{x^S[P(\partial,x)+(-1)^{S+1} P^*(\partial,-\partial-x)]\ ,
\quad P(\partial,x)\in\operatorname{Mat}_N\C[\partial,x]\} \ , \nonumber\\
\rmsk
&=& (I_{k,N}+(x+\partial)^S\bar{I}_{k,N})
\operatorname{Mat}_N\C[\partial,x] \ , \\
\rmxs
&=& \{(x+\partial)^S[P(\partial,x)+(-1)^{S+1} P^*(\partial,-\partial-x)]\ ,
\ P(\partial,x)\in\operatorname{Mat}_N\C[\partial,x]\} \ , \nonumber
\end{eqnarray}
where $I_{k,N}$ is defined in (\ref{ikn}) and $\bar{I}_{k,N}=\id-I_{k,N}$,
are subalgebras of $\gc_N$ for every $k=0,\dots,N$.
They are normalized with respect to the Virasoro element
$\lpms=\left(x+ \frac{1\mp S}{2}\partial\right)\id$.
\end{lemma}
\begin{proof}
The proof that $\rpmsk$ and $\rpmxs$ are subalgebras of $\gc_N$
is straightforward (see \cite{liberati}).
We need to prove that $\rpmsk,\ k=0,\dots,N$ and $\rpmxs$
are normalized subalgebras, namely:
\begin{eqnarray}\label{M8}
&& \lpms\ _{(i)}\ \rpmsk\ \subset\ \rpmsk \ , \\
&& \lpms\ _{(i)}\ \rpmxs\ \subset\ \rpmxs \ , \ \text{ for } i=0,1,2
\nonumber\ .
\end{eqnarray}
For $i=0,1$, (\ref{M8}) is obviously true and we are left to check
it for $i=2$.
By (\ref{xbracket}) one gets, after a straightforward computation:
\begin{eqnarray*}
&& \lps\ _{(2)}\ P(\partial,x)(I_{k,N}+x^S\bar{I}_{k,N}) \\
&& \quad =\frac{d^2}{d\lambda^2}\left.\left[
\left(x-\frac{S-1}{2}\partial\right)\id\ _\lambda\
P(\partial,x)(I_{k,N}+x^S\bar{I}_{k,N})
\right]\right|_{\lambda=0} \\
&& \quad =
\left\{x(2D_1-D_2)D_2+\partial D_1^2+SD_2+(2D_1-D_2)\right\}P(\partial,x) \\
&& \qquad \times (I_{k,N}+x^S\bar{I}_{k,N})
+2S(D_1-D_2)P(\partial,x)\cdot x^S\bar{I}_{k,N} \ ,
\end{eqnarray*}
where $D_1$ (resp. $D_2$) denotes partial derivative with respect to
$\partial$ (resp. $y$).
The left hand side is clearly in $\rpsk$ since, by (\ref{ikn}),
$I_{k,N}^2=I_{k,N},\ \bar{I}_{k,N}^2=\bar{I}_{k,N},\ I_{k,N}\bar{I}_{k,N}=0$,
so that we can write:
$$
x^S\bar{I}_{k,N}
=\bar{I}_{k,N}(I_{k,N}+x^S\bar{I}_{k,N})\ .
$$
This proves that $\rpsk$ is invariant under the action of $\lps\ _{(2)}$.
In the particular case $k=0$, the previous calculation gives:
$$
\lps\ _{(2)}\ x^S P(\partial,x)
=x^S \{x(2D_1-D_2)D_2+\partial D_1^2+(S+1)(2D_1-D_2)
P(\partial,x) \ .
$$
Since the differential operator in parenthesis is invariant under
the change of variables $\partial\rightarrow\partial,\
x\rightarrow-\partial-x$,
this equation clearly implies that $\rpxs$ in invariant
under the action of $\lps\ _{(2)}$.
A similar calculation shows that $\rmsk$ and $\rmxs$ are invariant
under the action of $\lms\ _{(2)}$.
\end{proof}
The particular choice of the matrix $I_{k,N}$ is not canonical,
and can always be redefined with a change of basis.
This fact is stated in the following:
\begin{lemma}
{\bfseries (i)} Suppose $P(x)\in\operatorname{Mat}_N\bC[x]$ is an invertible
matrix. Then we get an automorphism of $\gc_N$ by conjugation by $P(x)$:
$$
A(\partial,x)\mapsto P(\partial+x)A(\partial,x)P(x)^{-1} \ .
$$
{\bfseries (ii)} In particular, if $R$ is a subalgebra of $\gc_N$, so is
$R_P=P(\partial+x)RP(x)^{-1}$.
Moreover, if $R$ is normalized with respect to a Virasoro element $L$
of kind (\ref{virel}) and $P$ is a constant invertible matrix,
then $R_P$ is also normalized with respect to $L$.
The corresponding subalgebra of the reduced space is
$\pi(R_P)=P\pi(R)P^{-1}$.
\end{lemma}
\begin{proof}
The proof of (i) is straightforward and can be found in \cite{liberati}.
(ii) Follows immediately by the fact that $L$ is a scalar matrix, so that
it is unchanged after conjugation.
\end{proof}
We did not prove so far that all subalgebras $\rpmsk$ and
$\rpmxs\subset \gc_N$ act irreducibly on $\C[\partial]^N$.
In fact this is not true.
More precisely one can prove the following:
\begin{lemma}\label{irr}
{\bfseries (i)} $\rpsk$ and $\rpxs$ always act irreducibly
on $\C[\partial]^N$, \\
{\bfseries (ii)} $\rmsk$ and $\rmxs$ never act irreducibly
on $\C[\partial]^N$, unless $\rmsk=\gc_N$,
which happens only when $k=N$ or $S=0$,
or $\rmxs=\rpxs$, which happens only when $S=0$.
\end{lemma}
\begin{proof}
Since for any choice of the antiinvolution $*$ and
of the integer $k=0,\dots,n$
we have: $\rpmxs\subset\rpmsk$,
it will suffices to prove:
\begin{enumerate}
\item $\rpxs$ acts irreducibly on $\C[\partial]^N$,
\item there is a proper $\C[\partial]$--submodule $U_k\subset\C[\partial]^N$
which is invariant under the action of $\rmsk$, unless $k=N$ or $S=0$.
\end{enumerate}
Suppose $U\subset\C[\partial]^N$ is a non zero $\C[\partial]$--submodule
which is invariant under the $\lambda$--action of $\rpxs$.
Let us denote $U_0=U\cap \C^N$.
Let $u=\sum_{i=0}^k\partial^iv^i$ be any element of $U$ with
$v^i\in\C^N$ and $v^k\neq0$.
A generic element of $\rpxs$ is
$a=x^SP(\partial,x)$, where $P(\partial,x)$ is such that
$P(\partial,x)=(-1)^{S+1}P^*(\partial,-\partial-x)$.
In particular we can choose:
$P(\partial,x)=p_S(2x+\partial)\id$, where $p_S(y)=1$ for $S$ odd
and $p_S(y)=y$ for $S$ even.
By (\ref{xaction}) we have that the $\lambda$--action of $a$ on $u$ is:
$$
a_\lambda u=(\lambda+\partial)^Sp_S(\lambda+2\partial)
\sum_{i=0}^k(\lambda+\partial)^iv^i
\in U\otimes\C[\lambda] \ ,
$$
so that, if we look at the coefficient of the highest power of $\lambda$
(which is $S+k+$ deg($p_S$)), we get that $v^k\in U_0\neq0$.
Let us now consider elements:
$a=A,\ b=B(2x+\partial)\ \in \rpxs$, where $A\in V_{(-1)^{S+1}}$ and
$B\in V_{(-1)^S}$.
By taking the $\lambda$--action on a generic element $v\in U_0$ we get:
\begin{eqnarray*}
&& a_\lambda v=Av \quad \Rightarrow \quad Av\in U_0 \\
&& b_\lambda v=(\lambda+2\partial)Bv \quad \Rightarrow \quad Bv\in U_0 \ .
\end{eqnarray*}
But then $U_0$ is invariant under multiplication of both $V_+$ and $V_-$,
and since $V_++V_-=\operatorname{Mat}_N\C$, this is possible only
if $U_0=\C^N$.
Since $U$ by definition is a $\C[\partial]$--module and $\C^N\subset U$,
we finally get $U=\C[\partial]^N$.
We are left to prove the second part.
A generic element of $\rmsk$ is:
$a=(I_{k,N}+(x+\partial)^S\bar{I}_{k,N})P(\partial,x)$, where
$P(\partial,x)\in\operatorname{Mat}_N\bC[\partial,x]$.
Its action on $v(\partial)\in\C[\partial]^N$ is, by (\ref{xaction}):
$$
a_\lambda v(\partial)
=(I_{k,N}+\partial^S\bar{I}_{k,N})
P(-\lambda,\lambda+\partial)v(\lambda+\partial) \ .
$$
From this expression it is clear that:
$$
U=\left\{
\left(\begin{array}{c}
v_k(\partial) \\
\partial^Sv_{N-k}(\partial)
\end{array}\right)
\ , \
v_k(\partial)\in\C[\partial]^k\ ,\
v_{N-k}(\partial)\in\C[\partial]^{N-k}\right\} \ ,
$$
is a $\C[\partial]$--submodule which is invariant under
the action of $\rmsk$. Obviously $U$ is a proper submodule
as soon as $k\neq N$ and $S\neq0$.
\end{proof}
\begin{corollary}\label{wailin2}
{\bfseries (i)} Given $\lpms=\left(x+\frac{1\mp S}{2}\partial\right)\id$,
all the spaces $\hrpmsk$ defined by (\ref{rk}) are subalgebras
of the reduced space $\hgc_N$.
The corresponding normalized subalgebras of $\gc_N$ are:
$$
\C[\partial]\hrpmsk=\rpmsk
$$
{\bfseries (ii)} Given a left or right ideal $J\subset\operatorname{Mat}_N\C$,
depending whether $\sigma=+S$ or $-S$, the space $\hrpmsj$ is also
a subalgebra of $\hgc_N$, and it is obtained by $\hrpmsk$ for some
$k=0,\dots,N$ by conjugation by an invertible matrix:
$$
\hrpmsj=P\hrpmsk P^{-1} \ ,
$$
so that the corresponding normalized subalgebra of $\gc_N$ is:
$$
\C[\partial]\hrpmsj=P\rpmsk P^{-1} \ .
$$
{\bfseries (iii)} The space $\hrpmxs$ defined in (\ref{rx}), where
$*$ is a linear antiinvolution of $\operatorname{Mat}_N\C$,
is a subalgebra of the reduced space $\hgc_N$.
The corresponding normalized subalgebra of $\gc_N$ is:
$$
\C[\partial]\hrpmxs=\rpmxs \ .
$$
\end{corollary}
\begin{remark}
According to Corollary \ref{kari} there are no other subalgebras
$\hat{R}\subset\hgc_N$ with the property that $\sum_{n\geq0}V_n$
(defined by (\ref{1})) acts irreducibly on $\C^N$.
Thus, by Corollary \ref{baslem}, Lemma \ref{L} and Lemma \ref{irr},
we get that $\rpsk,\ k=0,\dots,N$ and $\rpxs$, where $*$ is an
antiinvolution of $\operatorname{Mat}_N\C$, are, up to conjugation
by an invertible constant matrix, all possibilities
of infinite rank subalgebras of $\gc_N$ acting irreducibly
on $\C[\partial]^N$.
\end{remark}
\begin{proof}
Consider for simplicity $\sigma=+S$; the argument for $\sigma=-S$ is the same,
after replacing left ideals with right ideals.
By definition:
\begin{eqnarray*}
\hrpsk=
&=& \left(\bigoplus_{0\leq n<S}Q^{(S)}_n(\partial,x)
\operatorname{Mat}_N\bC I_{k,N}\right)
\oplus\left(\bigoplus_{n\geq S}Q^{(S)}_n(\partial,x)
\operatorname{Mat}_N\bC\right) \\
&=& \hgc_NI_{k,N}+\hrps\otimes\operatorname{Mat}_N\bC \ ,
\end{eqnarray*}
where $\hrps=\bigoplus_{n\geq S}\C Q^{(S)}_n(\partial,x)\subset\gc_1$
was defined in Lemma \ref{remark4}.
It follows that the $\C[\partial]$--module generated by $\hrpsk$ is:
$$
\C[\partial]\hrpsk
= \gc_N I_{k,N}+\gc_N x^S \ .
$$
Here we used the fact that $\C[\partial]\hrps=x^S\gc_1$, as stated in
Corollary \ref{novembre}.
In order to prove (i), we just need to show that:
$$
\gc_NI_{k,N}+\gc_Nx^S=\gc_N(I_{k,N}+\bar{I}_{k,N}x^S) \ ,
$$
and this follows by the fact that $I_{k,N}$ and $\bar{I}_{k,N}$ are
complementary idempotent matrices: $I_{k,N}^2=I_{k,N}=\id-\bar{I}_{k,N}$.
We just proved $\hrpsk=\pi(\rpsk)$, which implies, by Corollary \ref{baslem}
and Lemma \ref{wailin}, that $\hrpsk$ is a subalgebra of $\hgc_N$.
For (ii), suppose the left ideal $J\subset\operatorname{Mat}_N\bC$ is
generated by the matrix $A$, namely $J=(\operatorname{Mat}_N\bC) A$.
If rk$(A)=k$, we can always find invertible matrices $P,\ Q$ such that
$A=QI_{k,N}P^{-1}$, which implies:
$$
J=P(\operatorname{Mat}_N\bC\ I_{k,N})P^{-1}.
$$
We then get, as we wanted, $\hrpsj=P\hrpsk P^{-1}$.
We are left to prove (iii).
By Lemma \ref{wailin}, $\rpxs$ is a subalgebra
of $\gc_N$ normalized with respect to $\lps$.
It follows by Corollary \ref{baslem} that the projection
$\hat{R}=\pi(\rpxs)$ is a subalgebra of the reduced space
$\hgc_N$.
We want to prove that, if we decompose $\hat{R}$ as in (\ref{1}),
then $V=\sum_{n\geq0}V_n\subset\operatorname{Mat}_N\C$ acts irreducibly
on $\C^N$.

By (\ref{qn}) we know that
$Q_S^{(S)}(\partial,x)=x^S$
and
$Q_{S+1}^{(S)}(\partial,x)=x^S(x+\frac{1}{2}\partial)$.
We then have, by definition of $\rpxs$, that
$Q_S^{(S)}(\partial,x)A,\ Q_{S+1}^{(S)}(\partial,x)B\ \in\rpxs$
for every $A\in V_{(-1)^{S+1}}$ and $B\in V_{(-1)^{S+2}}$,
or equivalently: $V_{(-1)^{S+1}}\subset V_S$
and $V_{(-1)^{S+2}}\subset V_{S+1}$.
But then $\operatorname{Mat}_N\C=V_++V_-\subset V$ and $V$ acts
irreducibly on $\C^N$.
Corollary \ref{kari} provides a list of all possibilities of
infinite--dimensional subalgebras $\hat{R}\subset\hgc_N$ such that
$\sum_{n\geq0}V_n$ acts irreducibly on $\C^N$.
It is not hard to understand, from this, that the only allowed possibility
is $\pi(\rpxs)=\hrpxs$, or equivalently
$\rpxs=\C[\partial]\hrpxs$, which is what we wanted.
\end{proof}


We thus obtained our main result:
\begin{theorem}\label{finthe}
{\bf (i)} Consider a Virasoro element in $\gc_N$ of the form
\begin{equation}\label{M9}
L=\left(x+\frac{1-\sigma}{2}\partial\right)\id \ .
\end{equation}
Let $\hgc_N=\bigoplus_{n\in\Z_+}
Q_n^{(\sigma)}(\partial,x)\operatorname{Mat}_N\bC$ be the
corresponding reduced space.
A complete list of subalgebras
$\hat{R}=\bigoplus_{n\in\Z_+}Q_n^{(\sigma)}(\partial,x)V_n\subset\hgc_N$
which are infinite--dimensional and such that
$\sum_{n\in\Z_+}V_n$ acts irreducibly on $\C^N$ is the following:
\begin{enumerate}
\item for $\sigma\notin\Z$ the only possibility is $\hat{R}=\hgc_N$;
\item for $\sigma\in\Z$, let $\sigma=\pm S$ with $S\in\Z_+$.
Then the only possibilities are:
$$
\hrpmsj
 = \left(\bigoplus_{0\leq n<S}Q_n^{(\sigma)}(\partial,x)J\right)
\oplus\left(\bigoplus_{n\geq S}Q_n^{(\sigma)}(\partial,x)
\operatorname{Mat}_N\bC\right) \ ,
$$
where $J$ is a left (resp. right) ideal of $\operatorname{Mat}_N\bC$ for
$\sigma=+S$ (resp. $\sigma=-S$), and
$$
\hrpmxs
 = \bigoplus_{n\geq S}Q_n^{(\sigma)}(\partial,x) V_{(-1)^{n+1}} \ ,
$$
where $V_\pm=\{A\in\operatorname{Mat}_N\bC\ |\ A^*=\pm A\}$ and $*$ is
a linear antiinvolution of $\operatorname{Mat}_N\bC$.
\end{enumerate}
{\bfseries (ii)} A complete list of infinite rank, normalized
(with respect to some Virasoro element of the form (\ref{M9})),
subalgebras of $\gc_N$ acting irreducibly on $\C[\partial]^N$
is the following:
$$
\rpsk
= \operatorname{Mat}_N\bC[\partial,x](I_{k,N}+x^S\bar{I}_{k,N}) \\
$$
and their conjugates by a constant invertible matrix,
$$
\rpxs
= \{x^S[P(\partial,x)+(-1)^{S+1} P^*(\partial,-\partial-x)] \ ,
\quad P(\partial,x)\in\operatorname{Mat}_N\C[\partial,x]\} \ ,
$$
where $S\in\Z_+$, $k=0,\dots,N$, $I_{k,N}$ is
defined by (\ref{ikn}), $\bar{I}_{k,N}=\id-I_{k,N}$
and $*$ is a linear antiinvolution of $\operatorname{Mat}_N\bC$.
The corresponding Virasoro element is
$\lps=\left(x+\frac{1-S}{2}\partial\right)\id$.
\end{theorem}

\ssubsection{An application to Jacobi polynomials $\dots$}

By Corollary \ref{wailin2} we have
$\hrpmxs\subset\rpmxs$, which means, in terms of basis elements,
that $Q_n^{(\pm S)}(\partial,x) A \in \rpmxs$ for
every $A\in V_{(-1)^{n+1}}$, $n\geq S$, or equivalently, by
(\ref{wa5}):
$$
Q_n^{(S)}(\partial,x) A
= x^S\tilde{Q}_n^{(S)}(\partial,x)A \ ,
$$
where:
$$
\tilde{Q}_n^{(S)}(\partial,x)A
=(-1)^{S+1}\tilde{Q}_n^{(S)}(\partial,-\partial-x)A^*
=(-1)^{n-S}\tilde{Q}_n^{(S)}(\partial,-\partial-x)A \ ,
$$
so that $\tilde{Q}_n^{(S)}(\partial,x)=
(-1)^{n-S}\tilde{Q}_n^{(S)}(\partial,-\partial-x)$. Similarly
$Q_n^{(-S)}(\partial,x)=(\partial+x)^S\tilde{Q}_n^{(-S)}(\partial,x)$,
with $\tilde{Q}_n^{(-S)}(\partial,x)=
(-1)^{n-S}\tilde{Q}_n^{(-S)}(\partial,-\partial-x)$.
Moreover it follows, by symmetry relation
(\ref{sym}) and the fact that
$Q_n^{(\pm S)}(\partial,y)$ is homogeneous of degree $n$, that:
\begin{eqnarray*}
\tilde{Q}_n^{(+S)}(\partial,x)
&=& (-1)^{n-S}\tilde{Q}_n^{(+S)}(\partial,-\partial-x)
 = (-1)^{n-S}\frac{Q_n^{(+S)}(\partial,-\partial-x)}{(-\partial-x)^S} \\
&=& (-1)^n\frac{Q_n^{(-S)}(-\partial,-x)}{(\partial+x)^S}
 = \frac{Q_n^{(-S)}(\partial,x)}{(\partial+x)^S}
 = \tilde{Q}_n^{(-S)}(\partial,x)
\end{eqnarray*}
Thus, we proved the following:
\begin{lemma}\label{ciao}
Let $S\in\Z_+$. The polynomials $Q_n^{(\pm S)}(\partial,x)$ defined
in (\ref{qn}) have the following properties:
\begin{eqnarray*}
Q_n^{(S)}(\partial,x)&=& x^S \tilde{Q}_n^{(S)}(\partial,x) , \ \\
Q_n^{(-S)}(\partial,x)&=& (x+\partial)^S \tilde{Q}_n^{(S)}(\partial,x) \ ,
\end{eqnarray*}
where $\tilde{Q}_n^{(S)}(\partial,x)$ is a homogeneous polynomial
in $\partial$ and $x$ of degree $n-S$ satisfying:
$$
\tilde{Q}_n^{(S)}(\partial,x)=\tilde{Q}_n^{(S)}(-\partial,\partial+x) \ .
$$
\end{lemma}
\begin{remark}
Equivalently, in terms of the $y$ variable defined in (\ref{yvar}),
the polynomials $R_n^{(\pm S)}(\partial,y)$ defined in (\ref{rn})
are such that:
\begin{eqnarray*}
R_n^{(S)}(\partial,y)&=& (y-\partial)^S \tilde{R}_n^{(S)}(\partial,y) \ , \\
R_n^{(-S)}(\partial,y)&=& (y+\partial)^S \tilde{R}_n^{(S)}(\partial,y) \ ,
\end{eqnarray*}
where $\tilde{R}_n^{(S)}(\partial,y)$ is a homogeneous polynomial
in $\partial$ and $y$ of degree $n-S$ satisfying:
$$
\tilde{R}_n^{(S)}(\partial,y)=\tilde{R}_n^{(S)}(-\partial,y) \ .
$$
\end{remark}
We can translate this result in terms of Jacobi polynomials,
simply by using (\ref{r-p}) (namely by putting $\partial=1$ in the
above relations for $R_n^{(S)}(\partial,y)$).
\begin{corollary}\label{parity}
For $S\in\Z_+$ the Jacobi polynomial $P^{(-S,S)}_n(y)$ is divisible by
$(y-1)^S$ and the Jacobi polynomial $P^{(S,-S)}_n(y)$ is divisible by
$(y+1)^S$. The ratio polynomials coincide:
\begin{equation}\label{par1}
\tilde{P}^{(S)}_n(y):=\frac{P^{(-S,S)}_n(y)}{(y-1)^S}
=\frac{P^{(S,-S)}_n(y)}{(y+1)^S} \ .
\end{equation}
Furthermore $\tilde{P}^{(S)}_n(y)$ is a polynomial
of degree $n-S$ with the parity of $n-S$:
$$
\tilde{P}^{(S)}_n(y)=(-1)^{n-S}\tilde{P}^{(S)}_n(-y) \ .
$$
\end{corollary}
\begin{remark}
This result is a generalization of the classical parity property
$P^{(0,0)}_n(y)=(-1)^nP^{(0,0)}_n(-y)$ of Legendre polynomials.
\end{remark}
Of course one can prove Corollary \ref{parity} directly
from the definition of Jacobi polynomials, but the proof is rather involved.
\ssubsection{$\dots$ and to products of Jacobi polynomials}


\noindent
Recall the coefficients $d^{(\sigma)}_{m,n,k}$ are defined in
(\ref{mln2}) and (\ref{dmnk}) by expanding the product of
two polynomials $Q_m^{(-\sigma)}(\partial,x)$ and
$Q_n^{(\sigma)}(\partial,x)$ in powers of $x$:
$$
Q_m^{(-\sigma)}(\partial,x)Q_n^{(\sigma)}(\partial,x)
=\sum_{k=0}^{m+n}\frac{1}{k!}d^{(\sigma)}_{m,n,k}\partial^kx^{m+n-k} \ .
$$
We can rewrite this relation in terms of products of Jacobi polynomials,
or rather hypergeometric functions, using relation (\ref{qnpn}):
$$
P_m^{(\sigma,-\sigma)}(2x+1)P_n^{(-\sigma,\sigma)}(2x+1)
=\sum_{l=0}^{m+n}D(\sigma;\ m,n,l)x^l \ ,
$$
where we denoted
$D(\sigma;\ m,n,l)
= \frac{\binom{2m}{m}\binom{2n}{n}}{(m+n-l)!}d^{(\sigma)}_{m,n,m+n-l}$,
which means, by (\ref{dmnk}):
\begin{equation}\label{Dmnl}
D(\sigma;\ m,n,l)
= \sum_{i,j:\atop{0\leq i\leq m
   \atop{0\leq j\leq n\atop{i+j=l}}}}
   \binom{m+i}{m}\binom{m+\sigma}{m-i}\binom{n+j}{n}\binom{n-\sigma}{n-j} \ .
\end{equation}
In this Section we will show how the classification of normalized subalgebras
of $gc_N$, namely Theorem \ref{finthe},
can be used to prove interesting properties
of the coefficients  $D(\sigma;\ m,n,l)$, whose direct prove would require
rather complicated identities involving sums of multinomial coefficients.


For any $m, n, l\in\Z_+$ such that $l\leq m+n$,
$D(\sigma;\ m,n,l)$ is by (\ref{Dmnl}) a polynomial in $\sigma$ of
degree less than or equal to $m+n-l$.
Furthermore it is manifestly invariant under
the simultaneous exchange $\sigma\leftrightarrow-\sigma$ and
$m\leftrightarrow n$, namely $D(\sigma;\ m,n,l)=D(-\sigma;\ n,m,l)$,
so that we can assume, without loss of generality,
$m\leq n$.


By Remark \ref{R} we have that if $\hat{R}=\bigoplus_{n\in\Z_+}X^nV_n$
is a subalgebra of the reduced space $\hgc_N$, then for
any $A\in V_m$ and $B\in V_n$ one has:
\begin{equation}\label{cond}
D(\sigma;\ m,n,l)AB +(-1)^{m+n-l+1} D(-\sigma;\ m,n,l)BA\ \in V_{l} \ .
\end{equation}
On the other hand Theorem \ref{finthe} provides us a list of
subalgebras of $\hgc_N$.
Let us use these two facts to get conditions on the coefficients
$D(\sigma;\ m,n,l)$.
For $\sigma=S\in \Z_+,\ k=0,\dots,N$, we have the subalgebra
$$
\hrpsk
= \left(\bigoplus_{0\leq n<S}X^n \operatorname{Mat}_N\bC\ I_{k,N}\right)\oplus
\left(\bigoplus_{n\geq S}X^n\operatorname{Mat}_N\bC\right) \ ,
$$
which means $V_n=\operatorname{Mat}_N\bC\ I_{k,N}$ for $n<S$
and $V_n=\operatorname{Mat}_N\bC$ for $n\geq S$.
For $m<S,\ n\geq S$ ans $l<S$, relation (\ref{cond}) becomes:
$$
D(S;\ m,n,l)AI_{k,N}B \pm D(-S;\ m,n,l)BAI_{k,N}\
\in \operatorname{Mat}_N\C\ I_{k,N} \ ,
$$
for every $A,B\in\operatorname{Mat}_N\C$, which of course implies
$D(S;\ m,n,l)=0$.
For $m,n\geq S$ and $l<S$, (\ref{cond}) becomes:
$$
D(S;\ m,n,l)AB +(-1)^{m+n-l+1} D(-S;\ m,n,l)BA\
\in \operatorname{Mat}_N\C\ I_{k,N} \ ,
$$
for every $A,B\in\operatorname{Mat}_N\C$,
which implies $D(S;\ m,n,l)=D(-S;\ m,n,l)=0$.
Finally, given a linear antiinvolution $*$ of $\operatorname{Mat}_N\bC$,
we have the subalgebra $\hrpxs = \bigoplus_{n\geq S}X^n V_{(-1)^{n+1}}$.
By taking $m,n,l\geq S$ one gets by (\ref{cond}) that,
for $A\in V_{(-1)^{m+1}}$ and $B\in V_{(-1)^{n+1}}$:
$$
D(S;\ m,n,l)AB +(-1)^{m+n-l+1} D(-S;\ m,n,l)BA\
\in V_{(-1)^{l+1}} \ ,
$$
which implies $D(S;\ m,n,l)=D(-S;\ m,n,l)$.
We just proved the following:
\begin{lemma}\label{facts}
Let $m,n,l\in\Z_+$ be such that $m\leq n$ and consider the function
$D(\sigma;\ m,n,l)$ defined in (\ref{Dmnl}).
It is a polynomial in $\sigma$ of degree less than or equal to $m+n-l$
such that:
\begin{enumerate}
\item $D(\sigma;\ m,n,l)=D(-\sigma;\ m,n,l)$ for $\sigma=0,1,\cdots,m$,
\item if $l< n$, then $D(\sigma;\ m,n,l)=0$ for $\sigma=l+1,\cdots,n$.
\end{enumerate}
\end{lemma}
\begin{remark}
Notice that the second part follows immediately by the explicit
expression (\ref{Dmnl}) of $D(\sigma;\ m,n,l)$,
but to check the first part directly is non trivial.
\end{remark}
We now want to use the following:
\begin{lemma}\label{postnikov}
Let $P(\sigma)$ be a polynomial in $\sigma$ of degree $d$ such that:
\begin{equation}\label{s}
P(y_i)=P(-y_i)\ ,\ i=1,\dots,s \ ,
\end{equation}
for distinct positive number $y_1,\dots,y_s$, and
\begin{equation}\label{r}
P(x_j)=0\ ,\ j=1,\dots,t \ ,
\end{equation}
for distinct positive number $x_1,\dots,x_t$. \\
{\bfseries (i)} If $2s\geq d$, then $P(\sigma)$ is an even polynomial
in $\sigma$. \\
{\bfseries (ii)} If $s+t\geq d$, but $2s<d$, then $P(\sigma)$
is uniquely defined, up to scalar multiplication, by the above
conditions (\ref{s}) and (\ref{r}).
In this case it must be $s+t=d$.
\end{lemma}
\begin{proof}
In general, if a polynomial $P(\sigma)$ satisfies (\ref{s}),
we can always decompose it as:
$$
P(\sigma)=(\text{ even polynomial })
+\prod_{i=1}^s(\sigma^2-i^2)(\text{ odd polynomial }) \ .
$$
(i) follows immediately by this
decomposition.
Suppose now $2s<d,\ s+t\geq d$. We denote $\bar{d}=d$ (resp. =d-1)
if $d$ is odd (resp. even).
If we write $P(\sigma)=\sum_{i=1}^da_i\sigma^i$, then (\ref{s}) and (\ref{r})
give the following condition on the coefficients $a_i,\ i=1,\dots,d$:
\begin{eqnarray*}
&& a_0+a_1x_1+\cdots+a_dx_1^d=0 \\
&& \qquad \cdots \\
&& a_0+a_1x_t+\cdots+a_dx_t^d=0 \\
&& a_1y_1+a_3y_1^3+\cdots+a_{\bar{d}}y_1^{\bar{d}}=0 \\
&& \qquad \cdots \\
&& a_1y_s+a_3y_s^3+\cdots+a_{\bar{d}}y_s^{\bar{d}}=0 \ .
\end{eqnarray*}
In order to prove (ii) it thus suffices to prove that the matrix:
$$
M=\left[\begin{array}{cccccc}
1 & x_1 & x_1^2 & x_1^3 & \cdots & x_1^{d-1} \\
1 & x_2 & x_2^2 & x_2^3 & \cdots & x_2^{d-1} \\
\vdots & & \vdots & & \cdots & \vdots \\
1 & x_t & x_t^2 & x_t^3 & \cdots & x_t^{d-1} \\
0 & y_1 & 0 & y_1^3 & \cdots &  \\
\vdots & & \vdots & & \cdots & \\
0 & y_s & 0 & y_s^3 & \cdots &
\end{array}\right]
$$
has rank greater than or equal to $d-1$. Without loss of generality
we can assume $s+t=d$.
If we denote by $A$ the $t\times(s+t)$ matrix obtained by taking the upper
$t$ rows of $M$
and by $B$ the $s\times(s+t)$ matrix consisting of the lower $s$ rows of $M$,
we get, by the Laplace expansion:
$$
\text{ det } M
= (-1)^{\sum_{k=t+1}^{t+s}k}\sum_{|I|=s}(-1)^{\sum_{l=1}^s i_l}
\text{ det }A_{I^c}\ \text{ det }B_I \ ,
$$
where $I=(i_1,\dots,i_s)$ is a subset of $(1,\dots,s+t)$ with $s$ elements,
$I^c=(1,\dots,s+t)-I=(j_1,\dots,j_t)$
and $B_I$ (resp. $A_{I^c}$) is the $s\times s$ (resp. $t\times t$)
matrix obtained by taking columns $i_1,\dots,i_s$ (resp. $j_1,\dots,j_t$)
of $B$ (resp. $A$).
Since det $B_I=0$ as soon as one of the $i_l$'s is odd,
all the signs in the sum are positive:
$$
\text{ det } M
= \pm \sum_{|I|=s}\text{ det }A_{I^c}\ \text{ det }B_I \ .
$$
Finally we notice that, if we denote
$\lambda=(i_1/2-1,i_2/2-1,\dots,i_s/2-1)$ and $\mu=(j_1-1,\dots,j_t-1)$,
then the determinants of $A_{I^c}$ and $B_I$ can be written in terms of
Schur's polynomials as:
\begin{eqnarray*}
\text{ det } A_{I^c}
&=& \prod_{0\leq h<k\leq t}(x_k-x_h)\ S_\lambda(x_1,\dots,x_t) \\
\text{ det } B_{I}
&=& \prod_{0\leq l\leq s}y_l\prod_{0\leq h<k\leq s}(y_k^2-y_h^2)\
S_\mu(y_1^2,\dots,y_s^2) \ ,
\end{eqnarray*}
and they are never zero, under our assumptions on $x_i$'s and $y_j$'s.
\end{proof}
We now want to combine Lemma \ref{facts} and Lemma \ref{postnikov}.
Suppose first that $0\leq l\leq m\leq n$.
In this case Lemma \ref{facts} guarantees that:
$$
D(\sigma;\ m,n,l)
=\prod_{i=l+1}^m(\sigma^2-i^2)R(\sigma) \ ,
$$
where $R(\sigma)$ is a polynomial of degree less than or equal to
$n-m+l$ such that:
\begin{eqnarray}\label{dom1}
R(\sigma) &=& R(-\sigma)\ , \ \text{ for } \sigma=0\dots,l \nonumber\\
R(\sigma) &=& 0\ ,\ \text{ for } \sigma=m+1,\dots,n \ .
\end{eqnarray}
We can therefore apply Lemma \ref{postnikov} to conclude:
\begin{enumerate}
\item for $n-m\leq l\leq m+n$, $R(\sigma)$ is an even polynomial in $\sigma$,
\item for $l<n-m$, $R(\sigma)$ is uniquely defined, up
to scalar multiplication, by the conditions (\ref{dom1}).
\end{enumerate}
For $0\leq m\leq l\leq n$, Lemma \ref{facts} and Lemma \ref{postnikov}
imply:
\begin{enumerate}
\item if $n-m\leq l\leq m+n$, then $D(\sigma;\ m,n,l)$ is an even polynomial
in $\sigma$,
\item is $l<n-m$, then $D(\sigma;\ m,n,l)$ is uniquely defined, up to scalar
multiplication, by conditions (\ref{dom1}).
\end{enumerate}
Finally, for $l>n$ one immediately gets by Lemma \ref{postnikov} that
$D(\sigma;\ m,n,l)$ is an even polynomial in $\sigma$.
We can summarize these results in the following:
\begin{corollary}
{\bfseries (i)} For $n-m\leq l\leq m+n$ one can write:
$$
D(\sigma;\ m,n,l)
= \prod_{i\in \cA_{l,n}}(i^2-\sigma^2)\ R(\sigma) \ ,
$$
where $\cA_{l,n}=\{l+1,\dots,n\}$ if $l<n$
and $\cA_{l,n}=\emptyset$ if $l\geq n$,
and $R(\sigma)$ is an even polynomial in $\sigma$ of degree
less than or equal to $m+n-l-2|\cA_{l,n}|$. \\
{\bfseries (ii)} If $l<n-m$, then:
$$
D(\sigma;\ m,n,l)
= \prod_{i\in \cA_{l,m}}(i^2-\sigma^2)
\prod_{i\in \cA_{m,n}}(i-\sigma)\
R(\sigma) \ ,
$$
where $R(\sigma)$ is a polynomial of degree
$m+n-l-2|\cA_{l,m}|-|\cA_{m,n}|$, uniquely defined, up to scalar
multiplication, by the conditions:
$$
R(\sigma)=
\frac{\prod_{i\in \cA_{m,n}}(i+\sigma)}{\prod_{i\in \cA_{m,n}}(i-\sigma)}
R(-\sigma)\ ,\ \text{ for } i=0,\dots,\min\{m,l\} \ .
$$
\end{corollary}

%
%
%
%


\bigskip\bigskip\noindent
\normalsize\textbf{Acknowledgments}\par
\smallskip
\par\noindent
We thank R. Askey and M. Rahman for correspondence about special functions
and A. Postnikov for helpful discussions on symmetric polynomials.
The first author is particularly grateful to E. Rassart
for providing valuable insight on orthogonal polynomials
and hypergeometric functions.



\end{document}